%% file: MainPaper.tex
\documentclass[mnsc,nonblindrev]{informs3_nojournal}

\usepackage{tikz,mathrsfs}
\usepackage{pgf}
\usetikzlibrary{arrows,automata}
\usepackage{amsfonts,amssymb}
\tikzstyle{legend}=[rectangle, draw=black,fill=white, rounded corners, minimum width=1cm, minimum height=0.75cm]
\OneAndAHalfSpacedXII 
\usepackage{endnotes}
\usepackage{ mathrsfs }
\usepackage{subcaption}
\usepackage[normalem]{ulem}
\newcommand{\ObjP}{{\mathcal{P}}} 
\newcommand{\ObjM}{{\mathcal{MS}}}  
\newcommand{\ObjA}{{\mathcal{SL}}} 

\usepackage{eurosym}
\let\footnote=\endnote

\newcommand{\edit}[1]{\textcolor{black} {#1}}

\usepackage{natbib}
 \bibpunct[, ]{(}{)}{,}{a}{}{,}%
 \def\bibfont{\small}%
 \def\bibsep{\smallskipamount}%
 \def\bibhang{24pt}%
\DeclareMathAlphabet{\mathpzc}{OT1}{pzc}{m}{it}
\TheoremsNumberedThrough     
\ECRepeatTheorems

\EquationsNumberedThrough    

\MANUSCRIPTNO{}
\begin{document}

\RUNAUTHOR{Besbes, Elmachtoub and Sun}
\RUNTITLE{Static Pricing: Universal Guarantees for Reusable Resources} 
\TITLE{Static Pricing: Universal Guarantees for Reusable Resources}

\ARTICLEAUTHORS{%
\AUTHOR{Omar Besbes}
\AFF{Columbia Business School, New York, NY, 10027, \EMAIL{ob2105@gsb.columbia.edu}}
\AUTHOR{Adam N. Elmachtoub}
\AFF{Department of Industrial Engineering and Operations Research \& Data Science Institute, Columbia University, New York, NY, 10027, \EMAIL{adam@ieor.columbia.edu}}
\AUTHOR{Yunjie Sun}
\AFF{Department of Industrial Engineering and Operations Research, Columbia University, New York, NY, 10027, \EMAIL{ys2888@columbia.edu}}
} 
\ABSTRACT{We consider a fundamental pricing model in which a fixed number of units of a reusable resource are used to serve customers. Customers arrive to the system according to a stochastic process and upon arrival decide whether or not to purchase the service, depending on their willingness-to-pay and the current price. The service time during which the resource is used by the customer is stochastic and the firm may incur a service cost. This model represents various markets for reusable resources such as cloud computing, shared vehicles, rotable parts, and hotel rooms. 
In the present paper, we analyze this pricing problem when the firm attempts to maximize a weighted combination of three central metrics: profit, market share, and service level.  Under Poisson arrivals, exponential service times, and standard assumptions on the willingness-to-pay distribution,  we establish a series of results that characterize the performance of static pricing in such environments. 

In particular, while an optimal policy is fully dynamic in such a context, we prove that a static pricing policy simultaneously guarantees 78.9\% of the profit, market share, and service level from the optimal policy. Notably, this result holds for any service rate and number of units the firm operates. \edit{Our proof technique relies on a judicious construction of a static price that is derived directly from the optimal dynamic pricing policy.} In the special case where  there are two units and the induced demand  is linear, we also prove that the static policy guarantees 95.5\% of the profit from the optimal policy. Our numerical findings on a large testbed of instances suggest that the latter result is quite indicative of the profit obtained by the static pricing policy across all parameters.}
\maketitle
\input{1_Introduction.tex}
\input{2_Model.tex}

\input{3_Analysis.tex}
\input{4_MultiObj.tex}
\input{5_ProfitMax.tex}

\input{7_Conclusion.tex}

\bibliographystyle{ormsv080} 

\bibliography{PRRBib} 
\ECSwitch 
\ECHead{Electronic Companion}
\input{6_Numeric.tex}
 \input{8_Appendix.tex}

\end{document}

%% file: 1_Introduction.tex
\section{Introduction}\label{Intro}
In many service industries, the same resource to serve one customer can be used to serve future customers  once the initial service is completed.  This type of resource is commonly referred to as a \textit{reusable resource}. For instance, in the hotel or car rental industry, a fixed number of rooms or vehicles are available to accommodate customers. Each customer uses one of these resources for some number of days until check out or return, after which it is free to be used by another customer. In a related example, many offices, campuses, or apartment buildings offer a pool of bicycles or vehicles to be rented or shared by their members, and units are always returned to their origin after being used by a member. Another example of a reusable resource is cloud computing servers, which can be used by customers to complete jobs after which they become available for processing new jobs. Finally, another interesting example of a reusable resource arises in the repair industry for aircraft, trains, and other large machinery. Specifically, there is a class of spare parts that are known as \textit{rotable}, meaning that when they break, the customer exchanges the broken part for a working part with the repair agent. When the repair agent receives the broken part, it is ``utilized'' for some time as it is being repaired, after which it becomes available again to service potential future customers. \edit{This application was the direct motivation for this paper, and the details of an implementation of this model at an aircraft OEM can be found in \citet*{besbes2018pricing}.}

All of the examples above share several important features which we shall incorporate into our model. First, the number of units available of each resource is  fixed (over appropriate time horizons), as acquiring more capacity or units involves significant investments. Second, when a resource is used, the service time is generally stochastic and varies across customers. Third, customers are price-sensitive and in turn the demand rate in each of these applications can be controlled by the price (which can be a one-time fee or an hourly/daily fee to the customer). Fourth, there is a cost incurred by the service provider associated with the usage of a unit  (e.g., in terms of cleaning, maintenance, or repair). Finally, in each of these settings it is highly unusual for a customer to wait for service. That is, if all units of the resource are occupied, the customer is typically lost.

In all of the settings above, the seller may have multiple objectives. The \emph{profit rate} is clearly a fundamental objective for any service provider, but typically such providers also focus significantly on their  \emph{market share}  and  \emph{service level}, i.e., the probability that an arriving customer finds a resource available. The latter two metrics are driven by the long term objectives of maintaining a prominent position in the market and ensuring that consumers find the service reliable.

In such environments, an optimal policy  will be highly dynamic in general, adjusting its prices often, as a function of the supply conditions. The question the present paper aims to address is the following. \textit{What is the performance of a simple static pricing (one price) policy in such  environments?} \edit{This question has dual practical and theoretical motivations. On the  one hand, in practice,  dynamic pricing may require systemic changes if prices are typically published in a catalog upfront. Dynamic prices may also be undesirable due to the negative perception by customers. On the other hand, the existing literature in dynamic pricing has argued for particular objectives that static pricing yields near-optimal performance in large-scale systems (see literature review). How robust is such an insight for a combination of objectives and for arbitrary scales? (While the scale for cloud computing may be large, it is often small as well, e.g., rotable spare parts.)}   In particular, in the present paper, we \edit{derive the first} \textit{universal performance guarantees} for static pricing with respect to the profit, market share, and service level objectives, with an optimal dynamic pricing strategy serving as the benchmark. In particular, we aim to provide results on the strength of static pricing that hold across all possible parameter regimes and scales.

To that end, we anchor our analysis around the following prototypical model. A service provider manages a pool of a single type of reusable resource. The firm uses the reusable resources to deliver service to customers over an infinite horizon. Customers arrive according to a Poisson process in which the rate depends on the price set by the firm. We make the standard assumption that the revenue rate is concave in the arrival rate. Upon arrival, a customer seizes one unit of the resources for an exponentially distributed random amount of time and pays a fee (which could depend on the realized duration of usage or be fixed in advance). The unit of resource occupied by a customer becomes available to others after service completion. The firm may also incur some cost of service. \edit{Note that the underlying dynamics correspond to an Erlang loss model with state-dependent arrival rates (\cite{erlang1917solution}).} The goal of the firm is to decide on the optimal pricing policy to maximize a combination of three different objectives: profit rate, market share, and service level.


The main contributions of the present paper lie in deriving universal performance guarantees for static pricing and can be summarized as follows.
\begin{itemize}

\item  We establish that for any combination of the three objectives --  profit rate, market share, and service level -- there exists a static pricing policy which can achieve at least 78.9\% of the value of each objective under the optimal dynamic pricing policy. This result holds for \emph{any capacity size}, \emph{market size}, and \emph{service rate}.

\item We consider a special case where the service provider is a profit maximizer, there are two units of the reusable resource, and the demand rate is linear in the price. We prove in this case that the static policy achieves at least 95.5\% of the optimal profit from dynamic pricing. This result holds for \emph{any market size} and \emph{service rate}.  

\item \edit{Our proof technique relies the construction of a judicious static price that is based on the \textit{actual} optimal dynamic pricing policy in the following manner. The static price is set so that arrival rate of the static policy (when there is at least 1 unit) is equal to the expected arrival rate of the optimal dynamic policy (when there is at least 1 unit). Under this construction, we lower bound the performance gap in terms of the stock-in probabilities of said static policy and the optimal policy, and analyze this ratio using a change of variables. Note that this is in stark contrast to the classical approach of relying on a fluid (deterministic) relaxation to obtain pricing heuristics and performance bounds.}

\item We complement the theoretical results above with numerical experiments over a broad test bed. These illustrate that the performance of static pricing is in general even better. Furthermore, for profit maximization, we find the performance of static pricing is always above 97.5\% of that obtained by an optimal dynamic pricing policy, indicating the robustness of the insights derived beyond the exact conditions assumed in the theorems. 

\end{itemize}

To the best of our knowledge, these are the \textit{first} universal guarantees derived for static pricing for this class of problems. Furthermore, the bounds derived highlight the very high performance of static pricing.




\subsection{Literature review}

We next provide an overview of the literature on the effectiveness of static pricing policies in the context of perishable inventory, queuing systems, and reusable resources. We note that although a server in a queueing system is indeed a reusable resource, these systems typically allow for customers to wait for service. In contrast, the reusable resources literature assumes that customers are immediately lost if no units are available.

The dynamic pricing literature has had an extensive focus on the context of perishable resources, where there is a finite time horizon to consume a finite number of units of one or more resources (see \cite{den2015dynamic} for a recent survey). The seminal work  of \citet{gallego1994optimal} shows that if the revenue function is concave, a static pricing policy loses at most $1/(2\sqrt{\min\{C,\lambda^*t\}})$, where $C$ is the number of units and $\lambda^*t$ represents the expected number of sales under the myopic price.  The authors also show a universal guarantee of $1-1/e$ for any parameter regime, with both results relying on a concavity assumption on the revenue rate (see also  \cite{ma2018dynamic}). \cite{ma2018dynamic} recently generalize these results for the same model without the concavity assumption, and also provide non-adaptive pricing policies for assortment optimization and non-stationary demand settings with constant factor performance guarantees. \cite{chen2018efficacy} showed that the $1-1/e$ guarantee and asymptotic optimality for static pricing also holds in the presence of strategic customers. \cite{gallego2008strategic} establish conditions for when static pricing is optimal in the presence of strategic customers. The value of static over dynamic pricing policies has also been considered in models with inventory cost and replenishment considerations, such as those in \cite{federgruen1999combined,chen2006optimal,yin2007joint, chen2010benefit}.  Related to static pricing policies are policies with limited price changes, such as those considered in \cite{feng1995optimal, bitran1997periodic, netessine2006dynamic, ccelik2009revenue, chen2015real, cheung2017dynamic}. Note that in our model there are no inventory costs, and inventory can be repeatedly reused over an infinite time horizon.

There is also an extensive literature on dynamic pricing in queues. 
\cite{paschalidis2000congestion} provide numerical results showing the promise of static pricing in multi-class systems.  \citet{ata2006dynamic} studied the dynamic pricing of an M/M/1 service system where the objective is welfare maximization, and numerically show it can have significant gains over static pricing.  \citet{maglaras2005pricing} showed in a service system, the revenue generated by the fluid-optimal prices are near optimal when the capacity and market potential are both large.  In a related model where the objective is revenue maximization, with observable queues, \citet{kim2017value} quantify more precisely the asymptotic value of dynamic pricing in large systems and prove conditions under which a two price policy is almost as good as a dynamic pricing policy. \citet{banerjee2015pricing} provide a queueing analysis of a ride-share platform where the customers are modeled as servers, and show that a static price is asymptotically equal to a dynamic price policy for large-scale systems, although dynamic pricing is more robust to modeling error.   

Closest to our formulation is the work of \citet{gans2007pricing} who study dynamic pricing to maximize the expected profit for rentals. Their model considers discounted rewards with a discrete price ladder, although with multiple customer types. They show the near-optimality of static pricing in highly utilized rental systems where both the offered load and system capacity are large.   To the best of our knowledge, all of the previously mentioned results quantifying the gap between static pricing and dynamic pricing  hold asymptotically when the scale of the system is large. In contrast,  our results provide universal guarantees that do not rely on the scale of the system.
Recently, \edit{\cite{waserhole2016pricing} and \cite{banerjee2016pricing} consider a general network of a single type of resource where prices control the rates between nodes, and prove a guarantee of $C/(C+n-1)$ (for  multiple objectives in the latter paper) but zero service times, where $n$ is the number of nodes and $C$ is the number of units. In related models, \edit{\cite{balseiro2019dynamic} introduce and analyze a Lagrangian-based dynamic pricing policy in a supply-constrained large network regime, while \cite{kanoria2019near} introduce an algorithm that does not require base arrival rate information with guarantees for transient performance. With non-zero service times, as we consider in our paper, \cite{banerjee2016pricing} provide a looser guarantee only when $C$ is large enough. Our paper provides a guarantee for any number of units, but does not consider a general network. Note that for many applications, the regime of interest has ``small" $C$, emphasizing the need for universal guarantees.  For example, for expensive rotable spare parts, most of the parts have $C\leq 5$ units  \citep*{besbes2018pricing}.} }

Various related studies focus on dynamic heuristics, multiple types of reusable resources, and other levers beyond pricing. \citet{lei2018real} study the dynamic pricing problem in a setting with deterministic service times and describe policies that are asymptotically optimal in the regime where demand and resource capacity are both large. 
\edit{\cite*{doan2019pricing} study the pricing problem of reusable resources under ambiguous distributions of demand and service time and use robust deterministic approximation models to construct asymptotically optimal fixed-price policies.}

Variants of the assortment optimization problem have been considered in \cite*{rusmevichientong2017dynamic}, \cite*{owen2018price} and \citet{gong2019online} with various universal approximation guarantees. The results in the first two papers can be extended to allow for dynamic pricing with discrete price points. \cite*{iyengar2004exponential} and \citet{levi2010provably} use  linear programming approaches to design admission control policies for such systems, which is a special case of dynamic pricing where a resource is either priced at a nominal price or at infinity. Their admission control policies are asymptotically optimal, and \citet{levi2010provably} also provides a universal guarantee of $1/2$. \cite*{chen2017revenue}  consider a generalization of this model that permits advanced reservations.



\subsection{Organization}
The paper is organized as follows. In Section \ref{Model}, we describe the  model along with structural properties of the optimal policy. In Section \ref{MultiObj}, we prove the 78.9\% performance guarantee of static pricing under our multi-objective setting for any parameter range. We then refine our guarantee to 95.5\% in Section \ref{ProfitMax} for the special case of profit maximization with two units under linear demand. \edit{Section \ref{Conclusion} concludes our paper and offers future directions of research. In an online appendix, we provide supplementary numerical experiments in Section \ref{NumEx} and provide all proofs that are omitted from the main body in Section \ref{Proofs}.}

%% file: 2_Model.tex
\section{Model and Preliminaries}\label{Model}
In this section, we first describe a general model for pricing a reusable resource. We then describe the various performance objectives the service provider may use, followed by several important properties of the optimal dynamic pricing policy.

\subsection{Pricing Model for a Reusable Resource}
We consider a model in which a service provider has a fixed number of identical, non-perishable units of a reusable resource that are sold to price-sensitive customers. The total number of units of the reusable resource that the provider has is $C$. At any point in time, each unit of the resource is either available for sale or occupied. Note that an occupied unit can be interpreted as a customer using the unit in the cloud computing and ride sharing examples, or being repaired in the rotable spare parts example from Section \ref{Intro}.

Customers arrive to the system over time according to a Poisson process with rate $\Lambda>0$. Each customer has an i.i.d. willingness-to-pay drawn from  a valuation distribution $F$. When a customer arrives, the provider offers a unit at some price $p$, and a customer decides to purchase usage of the resource if their willingness to pay exceeds $p$. We denote  by $\lambda(p):=\Lambda \bar{F}(p)$ the effective arrival rate at price $p$. When a customer decides to purchase usage, one unit is then occupied for a random amount of time that follows an exponential distribution with mean $1/\mu$. We assume that the usage times are i.i.d. across customers and independent of the customer valuations.

While a unit is being occupied, the firm cannot sell that unit until it is returned to the system, i.e., a customer finishes using the unit or the provider finishes repairing the unit. The firm incurs a cost $c$ to serve one customer, which may be a cleaning, maintenance, or repair cost. Any customer that arrives when all units are occupied is lost, regardless of the current price being offered. This assumption is largely motivated by the fact that in most of our applications the customers are seeking immediate service, and would naturally seek out a competitor if the provider has no units available.


We assume that there is a one-to-one correspondence between prices and effective arrival rates so that $\lambda(p)$ has a unique inverse, denoted by $p(\lambda)$. Therefore, one can view the effective arrival rate $\lambda$ as the decision variable. The firm dynamically determines a target effective arrival rate $\lambda$ which can be realized with the corresponding price $p(\lambda)$. From an analysis perspective, the effective arrival rate is more convenient to work with. We shall make the standard assumption in the revenue management literature that the profit rate function $\lambda (p(\lambda)-c)$ is concave in $\lambda$.

The set of admissible  policies, $\mathbf{\Pi}$, is the set of non-anticipating policies, i.e., policies such that the effective arrival rate at time $t$, $\lambda(t)$, may only depend on events up to $t^-$. We shall also be interested in the class of static policies, $\mathbf{\Pi}^s \subset \mathbf{\Pi}$, that simply fix a single arrival rate $\lambda$ (price) at every time $t$.



\subsection{Performance metrics}




One natural metric when selling the reusable resources is the expected profit rate. Fix a pricing policy $\pi$ and let $\lambda(t)$ denote the corresponding effective arrival rate at time $t$. Let $N^{\pi}(t)$ denote the corresponding arrival process. Note that the latter is a  non-stationary Poisson process with intensity $\lambda(t)$. Let $Q^{\pi}(t)$ denote the number of on-hand units at time $t$. The long-run average profit rate $\ObjP^{\pi}$ is given by
\begin{align}\label{Rpi}
\ObjP^{\pi} &= \liminf_{T \rightarrow \infty} \frac{1}{T} \mathbb{E}\left[ \int_0^T \mathbf{1}\{Q^{\pi}(t)>0\} (p(\lambda(t))-c)  dN^{\pi}(t) \right].
\end{align}
For simplicity in the exposition of the paper, we assume $p(\lambda(t))$ is a one-time fee a user pays for the service. Note that the analysis presented easily generalizes to the case when a user's payment depends on usage time, i.e., it is equivalent to charge a one-time price  that is simply the price per time unit multiplied by the expected usage time.



While the firm wants to maximize its profit, it may also want to keep a certain level of market share, i.e., the expected number of units sold, as well as a certain service level. The market share objective $\ObjM^{\pi}$ is directly aligned with maximizing sales, while the service level objective $\ObjA^{\pi}$ is measured by the fraction of time at least one unit is available.  These two objectives can be represented as 
\begin{equation*}
\ObjM^{\pi} = \liminf_{T \rightarrow \infty} \frac{1}{T} \mathbb{E}\left[  N^{\pi}(T) \right] 
\end{equation*}
and
\begin{equation*}
\ObjA^{\pi} = \liminf_{T \rightarrow \infty} \frac{1}{T} \mathbb{E}\left[ \int_0^T \mathbf{1}\{Q^{\pi}(t)>0\} dt \right]. 
\end{equation*}
Note that there is a trade-off between the various metrics; the optimal solution for one objective will generally be sub-optimal for another. For instance, maximizing the service level corresponds to setting a static price as large as possible, while maximizing market share corresponds to setting a static price of zero. Clearly neither price will result in any profit at all.

In order to take the different objectives into account simultaneously, we assume the firm maximizes a weighted combination of the objectives,
\begin{align}
\alpha_1 \ObjP^{\pi} + \alpha_2 \ObjM^{\pi} + \alpha_3 \ObjA^{\pi},
\end{align}
where $\alpha_1, \alpha_2, \alpha_3 \geq 0$ are the weights placed on each objective by the service provider. Without loss of generality, we assume  $\alpha_1 + \alpha_2 + \alpha_3  = 1$. We let $V^*$ denote the long-run value under the optimal policy, and is thus defined by 
\begin{align}
V^* \ := \ &\sup_{\pi \in {\mathbf{\Pi}}} \ \left\{ \alpha_1 \ObjP^{\pi} + \alpha_2 \ObjM^{\pi} + \alpha_3 \ObjA^{\pi} \right\}. \label{Eq1}
\end{align}
We denote by $\pi^*$ an optimal policy.  Similarly, we let $V^s$ denote the long-run value under the optimal static policy, and is thus defined by 
\begin{align}
V^{s} \ := \ &\sup_{\pi \in {\mathbf{\Pi}^s}} \ \left\{ \alpha_1 \ObjP^{\pi} + \alpha_2 \ObjM^{\pi} + \alpha_3 \ObjA^{\pi} \right\}. \label{eq:static_obj}
\end{align}

In the present paper, we focus on universal performance guarantees for static pricing. In particular, we shall focus on the worst-case performance of the optimal static pricing policy in comparison to the optimal dynamic policy. That is, we seek to characterize the maximum possible loss over all possible instances of our model. Formally,  we let $\Omega$ denote the family of instances 
\begin{equation*}
\Omega := \{(C,\mu,p(\cdot),c,\alpha_1,\alpha_2,\alpha_3): C\in \mathbb{N}^{+}, c,\mu>0, \alpha_1+\alpha_2+\alpha_3=1, \alpha_i\geq 0, \lambda (p(\lambda)-c) \text{ concave in }  \lambda\}.
\end{equation*}
In turn, we aim to provide a universal lower bound on
\begin{equation*}
\inf_{\Omega} \frac{V^{s}}{V^*},
\end{equation*}
which is the ratio between the objectives under the optimal static and dynamic pricing policies. In fact, we shall show that our bound applies to the corresponding ratios of each of the three objectives.

%% file: 3_Analysis.tex
\subsection{Analysis of the benchmark $V^*$}\label{Analysis}
We shall now characterize the structure of an optimal solution to the dynamic pricing problem stated in Equation \eqref{Eq1}. Given the Poisson assumption on arrivals and the exponential assumption on service times, without loss of optimality, one may focus on  stationary policies that update the price only at changes in the number of units on-hand. The memoryless property of the exponential distribution allows us to fully characterize the system (a continuous-time Markov chain) by the number of units on-hand.
As we shall see, this allows us to provide closed-form expressions for the steady state distribution and objectives under a particular policy.


An admissible policy $\pi$ may be represented by $C$ arrival rates $\lambda_1, \ldots \lambda_C$. When the provider has only $i$ units available, the price is set to $p(\lambda_i)$. Note that the static policy is a special case where $\lambda_1=\ldots =\lambda_C$. Furthermore, the system can now be modeled as a birth-death process where each state represents the number of units available. The transition rate from state $i$ to $i+1$ is $(C-i) \mu$ for $i=0,\ldots, C-1$. The transition rate from $i$ to $i-1$ is $\lambda_i$ for $i=1,\ldots, C$. A standard calculation for computing the steady state probabilities, $\mathbb{P}_i(\pi)$ yields that
\begin{align*}
\mathbb{P}_i(\pi) 
&=\frac{C!}{(C-i)!}\frac{\Pi_{j=i+1}^C \frac{\lambda_j}{\mu}}{\sum_{k=0}^C \frac{C!}{(C-k)!}\Pi_{j=k+1}^C\frac{\lambda_j}{\mu}}, \qquad i=0,\ldots,C.
\end{align*}

Using the steady state probabilities, we may express our three objectives simply as 
\begin{align}
\ObjP^{\pi} & =  \sum_{i=1}^C \lambda_i (p(\lambda_i)-c) \mathbb{P}_i(\pi) \label{eq:profit} \\ 
\ObjM^{\pi} & =  \sum_{i=1}^C \lambda_i \mathbb{P}_i(\pi), \label{eq:marketshare}\\
\ObjA^{\pi} & =\sum_{i=1}^C  \mathbb{P}_i(\pi)= 1-\mathbb{P}_0(\pi).  \label{Eq6}
\end{align}

Let us denote by $\lambda_i^*$  the effective arrival rate in state $i$ under the optimal policy, and by $\mathbb{P}^*_i$   the steady-state probabilities of being in state $i$ under the optimal policy. In Lemma \ref{lemma Structural}, we show a fundamental property that effective arrival rates are decreasing as the number of units available increases. Moreover, all such arrival rates do not exceed the myopic rate $\bar{\lambda}$ (the rate only maximizes the immediate reward without considering the future), which yields the highest possible instantaneous objective rate.
\begin{lemma}\label{lemma Structural}
Let $\lambda_i^\ast$ be the optimal arrival rate when the on-hand inventory level is $i$. Let $\bar{\lambda}$ denote the myopic arrival rate where $\bar{\lambda}=\arg\max_\lambda \lambda(\alpha_1(p(\lambda)-c)+\alpha_2)$. Then
	\begin{equation}\label{Eq5}
	\bar{\lambda}\ \geq \ \lambda_C^\ast \ \geq \ \cdots \ \geq  \ \lambda_1^\ast.
	\end{equation}
\end{lemma}

The proof of Lemma \ref{lemma Structural} can be found in Section \ref{Proofs}.
Notice that the result presented in Lemma \ref{lemma Structural} shares the same structural property as presented in Theorem 1 in \citet{gans2007pricing} where the objective is only profit maximization in a discounted reward setting. We extend the analysis to a long-run average reward setting with multiple objectives and prove that monotonicity of optimal prices (and rates) still holds. We will make use of this property in the subsequent analysis.

%% file: 4_MultiObj.tex
\section{Static Pricing Guarantee for Multi-Objective Optimization}\label{MultiObj}


We next investigate the performance of static pricing and present our first main result.

\begin{theorem}\label{Thm Single}
There exists a  static pricing policy $\pi^s$ that guarantees at least $\frac{15}{19}$ of the profit rate, market share, and service level of the optimal dynamic pricing policy. 
Equivalently,
\begin{equation*}
\inf_{\Omega} \min\left\{\frac{\ObjP^{\pi^s}}{\ObjP^{\pi^*}},\frac{\ObjM^{\pi^s}}{\ObjM^{\pi^*}},\frac{\ObjA^{\pi^s}}{\ObjA^{\pi^*}}\right\}\geq \frac{15}{19}.
\end{equation*}
\end{theorem}

Theorem \ref{Thm Single} provides a strong guarantee:  there exists a static price that nearly approximates the performance of an optimal dynamic pricing policy. Specifically, this price guarantees that the profit rate, market share, and service level are  at least $\frac{15}{19}\approx .789$ of the corresponding values under the dynamic pricing policy. Of course, a direct consequence of Theorem \ref{Thm Single} is that the optimal single price will have an overall objective of at least $0 .789$ of the objective under the optimal dynamic pricing policy as well. It is important to note that our result makes no assumption on the number of units in the system, demand rate, or service rate. This is in stark contrast to the previous literature which require the system usage and capacity to be large to provide theoretical guarantees. 

It is worthwhile to note that our proof is constructive and exhibits a particular static price that yields such performance. The static price behind our major finding is constructed using the optimal policy, which we denote by $\pi^*$. Recall that $\lambda_i^*$ are the arrival rates under the optimal policy and $\mathbb{P}^*_i$ are  the steady-state probabilities. The single price is simply chosen so that the corresponding arrival rate, $\tilde{\lambda}$, is the same as the expected arrival rate under the optimal policy when units are available. More specifically, the static arrival rate $\tilde{\lambda}$ is selected so that
\begin{align}
\tilde{\lambda}=\frac{\sum_{i=1}^C \lambda_i^*\mathbb{P}_i^*}{\sum_{i=1}^C \mathbb{P}_i^*}=\frac{\sum_{i=1}^C \lambda_i^*\mathbb{P}_i^*}{1-\mathbb{P}_0^*}. \label{eq:lambda}
\end{align}
\edit{In our proof, which we detail in the next subsection, we show that the performance guarantee of optimal static pricing can be lower bounded by the ratio of the stock-in probabilities of the specific static policy $\tilde{\lambda}$ and the optimal policy. Note that the stock-in probabilities of both policies can be expressed in terms of the optimal arrival rates $\lambda_i^*$, which allows us to focus on lower bounding a closed-form quantity (a ratio of high-dimensional polynomials). }




\subsection{Proof of Theorem \ref{Thm Single}} \label{sec:proof}
The proof is organized around two main steps. In the first step, we exploit the concavity of the revenue rate function (in the quantity space) to establish  that for each of the three objectives, the ratio of the performances under the static and optimal policies is at least the ratio of the corresponding service levels. The second step bounds the ratio of the service levels by $15/19$ by enumerating several cases, with each case proven using elementary calculus. A key component of this second step is a change of variables from demand rates to the product of demand rates. Both steps fundamentally exploit the explicit construction of $\tilde{\lambda}$ in Eq. \eqref{eq:lambda}.  With some abuse of notation, we index quantities with $\tilde{\lambda}$ to denote these under the static policy induced by this static rate.

\textbf{Step 1.} In the first step, we lower bound each of $\frac{\ObjP^{\tilde{\lambda}}}{\ObjP^{\pi^*}}$, $\frac{\ObjM^{\tilde{\lambda}}}{\ObjM^{\pi^*}}$, and $\frac{\ObjA^{\tilde{\lambda}}}{\ObjA^{\pi^*}}$ by $\frac{1-\mathbb{P}_0(\tilde{\lambda})}{1-\mathbb{P}_0^*}$. Note that by Eq. \eqref{Eq6}, the lower bound is exact for the service level ratio, i.e.,
\begin{align}
\frac{\ObjA^{\tilde{\lambda}}}{\ObjA^{\pi^*}}=\frac{1-\mathbb{P}_0(\tilde{\lambda})}{1-\mathbb{P}_0^*}. \label{eq:rA}
\end{align}
The lower bound is also exact for the market share ratio. Using Eqs. \eqref{eq:marketshare} and \eqref{eq:lambda}, we have that
\begin{align}
\frac{\ObjM^{\tilde{\lambda}}}{\ObjM^{\pi^*}}&=\frac{\tilde{\lambda}(1-\mathbb{P}_0(\tilde{\lambda}))}{\sum_{i=1}^C\lambda_i^*\mathbb{P}_i^*}=\frac{\frac{\sum_{i=1}^C \lambda_i^*\mathbb{P}_i^*}{1-\mathbb{P}_0^*}(1-\mathbb{P}_0(\tilde{\lambda}))}{\sum_{i=1}^C\lambda_i^*\mathbb{P}_i^*}=\frac{1-\mathbb{P}_0(\tilde{\lambda})}{1-\mathbb{P}_0^*}. \label{eq:rS}
\end{align}
For  the profit ratio by the ratio, we have
\begin{align}
\frac{\ObjP^{\tilde{\lambda}}}{\ObjP^{\pi^*}}
&\ = \ \frac{\tilde{\lambda}(p(\tilde{\lambda})-c)(1-\mathbb{P}_0(\tilde{\lambda}))}{  \sum_{i=1}^C \lambda_i^* (p(\lambda_i^*)-c) \mathbb{P}_i^* } \nonumber  \\
&\ = \ \frac{\tilde{\lambda}(p(\tilde{\lambda})-c)}{  \sum_{i=1}^C \lambda_i^* (p(\lambda_i^*)-c) \frac{\mathbb{P}_i^*}{1-\mathbb{P}_0^*}} \cdot \frac{1-\mathbb{P}_0(\tilde{\lambda})}{1-\mathbb{P}_0^*} \nonumber \\
&\ \geq \ \frac{\tilde{\lambda}(p(\tilde{\lambda})-c)}{\tilde{\lambda}(p(\tilde{\lambda})-c)} \cdot \frac{1-\mathbb{P}_0(\tilde{\lambda})}{1-\mathbb{P}_0^*} \nonumber \\
&\ = \ \frac{1-\mathbb{P}_0(\tilde{\lambda})}{1-\mathbb{P}_0^*}. \label{eq:rR}
\end{align}
The first equality follow from Eq. \eqref{eq:profit}. The inequality follows from the fact that the function $\lambda(p(\lambda)-c)$ is concave in $\lambda$ and applying Jensen's inequality to a random variable that takes value $\lambda_i^*$ with probability $\frac{\mathbb{P}_i^*}{1-\mathbb{P}_0^*}$ for $i=1,\ldots,C$. Note that the expected value of this random variable is exactly $\tilde{\lambda}$ by Eq. \eqref{eq:lambda}. We next characterize the stock-in probabilities, and the remainder of the proof, in terms of the new $z$ variables. This variable transformation unlocks the ability to apply (many) basic calculus ideas to prove our lower bound.


\textbf{Step 2.} To find the lower bound of the ratio of stock-in probabilities, we define a set of auxiliary notation which will be useful in our subsequent analysis. We define $a_i := \frac{C!}{(C-i)!}$ for $i=0,1,\ldots, C$ and $z_i:= \Pi_{j=i}^C \frac{\lambda_j^*}{\mu}$ for $i=1, \ldots, C+1$. For clarity, note that $z_{C+1}=1$. We also define $x := \sum_{k=1}^{C} a_kz_{k+1}$ and $y := \sum_{k=2}^{C} a_kz_k$. \edit{Our analysis first relies on a change of variables from the $lambda$-space to the $z$-space.}

Using the steady-state probabilities derived in Section \ref{Analysis} and the definition of $\tilde{\lambda}$, the service levels of the static and dynamic policies can be written as 
\begin{align*}
1-\mathbb{P}_0^* &= \frac{\sum_{i=1}^C a_iz_{i+1}}{\sum_{i=0}^C a_iz_{i+1}} \\
1-\mathbb{P}_0(\tilde{\lambda}) &= \frac{\sum_{i=1}^C a_i [(\sum_{k=1}^C a_kz_k)/(\sum_{k=1}^C a_kz_{k+1})]^{C-i}}{\sum_{i=0}^C a_i [(\sum_{k=1}^C a_kz_k)/(\sum_{k=1}^C a_kz_{k+1})]^{C-i}}.
\end{align*}

From the above, it is clear that the ratio of the service levels may be written as a function of $z_1,\ldots,z_C$. We call this function $R(z_1,\ldots,z_C)$. Formally, 
\begin{equation*}
R(z_1,\cdots,z_C):= \frac{1-\mathbb{P}_0(\tilde{\lambda})}{1-\mathbb{P}_0^*}=\frac{(\sum_{k=0}^{C} a_kz_{k+1})(\sum_{i=1}^C a_i [(\sum_{k=1}^C a_kz_k)/(\sum_{k=1}^C a_kz_{k+1})]^{C-i})}{(\sum_{k=1}^{C} a_kz_{k+1})(\sum_{i=0}^C a_i [(\sum_{k=1}^C a_kz_k)/(\sum_{k=1}^C a_kz_{k+1})]^{C-i})}.
\end{equation*}

We next develop a uniform lower bound on $R(z_1,\cdots,z_C)$ by developing separate bounds for the cases where $C$ is small ($C\le 3$) or large ($C \ge 4$).

\textbf{Step 2a.} We prove the lower bound for the cases where $C$ is at most 3. When $C=1$, then $\tilde{\lambda}=\lambda_1^*$ and therefore $R(z_1)=1$. When $C=2$, we have
\begin{equation*}
R(z_1,z_2) = \frac{z_1^{2}+4z_1z_2+3z_1+4z_2^{2}+6z_2+2}{z_1^{2}+4z_1z_2+2z_1+5z_2^{2}+6z_2+2}\geq \frac{4}{5},
\end{equation*}
where the inequality follows by matching terms and looking at the minimum ratio.
When $C=3$, the numerator of $R(z_1,z_2,z_3)$ is
\begin{align*}
&48+192z_3+120z_2+32z_1+264z_3^{2}+336z_2z_3+96z_1z_3+108z_2^{2}+64z_1z_2+10z_1^{2}+120z_3^{3}\\
&+228z_2z_3^{2}+68z_1z_3^{2}+144z_2^{2}z_3+88z_1z_2z_3+14z_1^{2}z_3+30z_2^{3}+28z_1z_2^{2}+9z_1^{2}z_2+z_1^{3}
\end{align*}
while the denominator of $R(z_1,z_2,z_3)$ is
\begin{align*}
&48+192z_3+120z_2+24z_1+264z_3^{2}+336z_2z_3+72z_1z_3+108z_2^{2}+48z_1z_2+6z_1^{2}+128z_3^{3}\\
&+252z_2z_3^{2}+60z_1z_3^{2}+168z_2^{2}z_3+84z_1z_2z_3+12z_1^{2}z_3+38z_2^{3}+30z_1z_2^{*2}+9z_1^{2}z_2+z_1^{3}.
\end{align*}
By matching terms in the numerator and denominator, it is then clear that
\begin{equation*}
R(z_1,z_2,z_3)\geq\frac{30}{38}=\frac{15}{19}.
\end{equation*}

\textbf{Step 2b.} Next, we consider the case where $C\geq 4$. While the function ratio $R(\cdot)$ is difficult to analyze directly, we will  derive a lower bound on $R$, which we denote by $\tilde{R}(\cdot)$, which will be amenable to analysis.  The bound can be derived simply by observing that
\begin{align*}
R(z_1,\ldots,z_C) &=\frac{(\sum_{k=0}^{C} a_kz_{k+1})(\sum_{i=1}^C a_i [(\sum_{k=1}^C a_kz_k)/(\sum_{k=1}^C a_kz_{k+1})]^{C-i})}{(\sum_{k=1}^{C} a_kz_{k+1})(\sum_{i=0}^C a_i [(\sum_{k=1}^C a_kz_k)/(\sum_{k=1}^C a_kz_{k+1})]^{C-i})}\\
&=\frac{(\sum_{k=0}^{C} a_kz_{k+1})[\sum_{i=1}^C a_i (\sum_{k=1}^{C} a_kz_k)^{C-i}(\sum_{k=1}^{C} a_kz_{k+1})^{i-1}]}{[\sum_{i=0}^C a_i (\sum_{k=1}^{C} a_kz_k)^{C-i}(\sum_{k=1}^{C} a_kz_{k+1})^{i}]}\\
&\geq\frac{(\sum_{k=0}^{C} a_kz_{k+1})[\sum_{i=1}^4 a_i (\sum_{k=1}^{C} a_kz_k)^{C-i}(\sum_{k=1}^{C} a_kz_{k+1})^{i-1}]}{[\sum_{i=0}^4 a_i (\sum_{k=1}^{C} a_kz_k)^{C-i}(\sum_{k=1}^{C} a_kz_{k+1})^{i}]}\\
&=\frac{(\sum_{k=0}^{C} a_kz_{k+1})[\sum_{i=1}^4 a_i (\sum_{k=1}^{C} a_kz_k)^{4-i}(\sum_{k=1}^{C} a_kz_{k+1})^{i-1}]}{[\sum_{i=0}^4 a_i (\sum_{k=1}^{C} a_kz_k)^{4-i}(\sum_{k=1}^{C} a_kz_{k+1})^{i}]} \\
&=: \tilde{R}(z_1,\ldots,z_C).
\end{align*}

Next, we derive a lower bound on $\tilde{R}(\cdot)$ though two subcases, depending on the ratio of $y$ to $z_1$.
We first establish in Lemma \ref{Derivative} (proved in  Section \ref{Proofs}) that the partial derivative with respect to the first argument is non-negative as long as $y\geq a_1z_1$.
\begin{lemma}\label{Derivative}
Fix  $C\geq 4$. Fix $z_1, z_2,...,z_C$ $\in$ $[0,\infty)^{C}$ and suppose $y\geq a_1z_1$, then
\begin{align*}
\frac{\partial \tilde{R}}{\partial z_1}\geq 0.
\end{align*}
\end{lemma}
When $y\geq a_1 z_1$, Lemma \ref{Derivative} implies that the worst case value of $\tilde{R}$ occurs when $z_1=0$.  In turn, in Lemma \ref{Cal bound} (proved in Section \ref{Proofs}), we establish a uniform lower bound on $\tilde{R}(0,z_2,\ldots,z_C)$. 
\begin{lemma}\label{Cal bound}
Fix $C\geq 4$. For all $z_2,...,z_C$ $\in$ $[0,\infty)^{C-1}$, we have  
\begin{equation*}
\tilde{R}(0,z_2,\ldots,z_C) \geq \frac{104}{131}.
\end{equation*}
\end{lemma}
From Lemmas \ref{Derivative} and \ref{Cal bound}, we can conclude that when $y\geq a_1 z_1$, then $\tilde{R}(z_1,z_2,\ldots,z_C) \geq \tilde{R}(0,z_2,\ldots,z_C) \geq \frac{104}{131}$.

If $y \leq a_1z_1$, then there is no guarantee on the derivative, but one may directly establish a uniform lower bound on $\tilde{R}$ as articulated in Lemma \ref{Cal bound2} (proved in Section \ref{Proofs}).
\begin{lemma}\label{Cal bound2}
Fix $C\geq 4$ and suppose $y \leq a_1 z_1$,  then
\begin{align*}
\tilde{R}(z_1,z_2,\ldots,z_C) \geq \frac{6}{7}.
\end{align*}
\end{lemma}

Combining both cases, We conclude that  $R(z_1,z_2,\ldots,z_C)\geq  \tilde{R}(z_1,z_2,\ldots,z_C) \geq \min \{\frac{104}{131}, \frac{6}{7} \} \geq \frac{15}{19}$. This completes the proof of Theorem \ref{Thm Single}.

\subsection{Tightness of analysis} \label{sec:tightness}
We present an example which shows that the lower bound of $\frac{15}{19}$ in Theorem \ref{Thm Single}  can be tight for a family of instances. That is, we shall describe instances in which the static policy we construct, $\tilde{\lambda}$, achieves exactly a fraction $15/19$ of the optimal dynamic policy. Namely, we shall fix $C=3$, $\alpha_1=0$, $\alpha_2=0$, $\alpha_3=1$, $\mu$ to be arbitrarily close to 0, and $p(\lambda)=\frac{1}{\lambda}$.

Since $\alpha_3=1$, then the objective is to maximize the service level, that is
\begin{equation*}
\max_\pi \ObjA^{\pi} = 1-\mathbb{P}(\pi).
\end{equation*}
The service level is always bounded above by 1,  and hence it is clear that the policy $(\lambda_1^*, \lambda_2^*, \lambda_3^*)=(0,\Lambda,\Lambda)$ is optimal since 
\begin{align*}
\ObjA^{(0,\Lambda,\Lambda)} = \frac{\sum_{i=1}^3\frac{6}{(3-i)!}\mu^i\Pi_{j=i+1}^3\lambda_j^*}{\sum_{i=0}^3\frac{6}{(3-i)!}\mu^i\Pi_{j=i+1}^3\lambda_j^*}
= \frac{\sum_{i=1}^3\frac{6}{(3-i)!}\mu^i\Pi_{j=i+1}^3\lambda_j^*}{0+\sum_{i=1}^3\frac{6}{(3-i)!}\mu^i\Pi_{j=i+1}^3\lambda_j^*}
= 1.
\end{align*}

Now let us consider the static policy $\tilde{\lambda}$ which we construct according to Eq. \eqref{eq:lambda}. Recall from Section \ref{sec:proof} that the performance of the static pricing policy with respect to the service level and market share objectives is $R(z_1,z_2,z_3)$, where 
$z_i:= \Pi_{j=i}^C \frac{\lambda_j^*}{\mu}$. In addition, the performance of the static pricing policy with respect to the profit rate is also $R(z_1,z_2,z_3)$ because $\lambda(p(\lambda)-c)$ is linear in $\lambda$ if $p(\lambda)=\frac{1}{\lambda}$, which makes the Jensen's inequality tight in the derivation of Eq. \eqref{eq:rR}. Since $z_1=0$, then  the ratio becomes 
\begin{align*}
&R(z_1,z_2,z_3) 
\\&= \frac{48+192z_3+120z_2+264z_3^{2}+336z_2z_3+108z_2^{2}+120z_3^{3}+228z_2z_3^{2}+144z_2^{2}z_3+30z_2^{3}}{48+192z_3+120z_2+264z_3^{2}+336z_2z_3+108z_2^{2}+128z_3^{3}+252z_2z_3^{2}+168z_2^{2}z_3+38z_2^{3}}.
\end{align*}
Since $z_2 = \frac{\Lambda^2}{\mu^2}$ and $z_3=\frac{\Lambda}{\mu}$,  we have $z_3 \rightarrow \infty$ and $z_3 =o(z_2)$ as $\mu \rightarrow 0$, and hence
\begin{equation*}
\lim_{\mu \to 0} R(z_1,z_2,z_3) = \frac{30}{38} = \frac{15}{19}.
\end{equation*}

%% file: 5_ProfitMax.tex
\section{Sharpening the Bound for Profit Maximization }\label{ProfitMax}


In this section, we seek to focus more deeply on the profit maximization objective corresponding to $\alpha_1=1$. This objective is central in the literature and we aim to understand to what extent can our $78.9\%$ guarantee from Section \ref{MultiObj} be improved. 

\begin{theorem}\label{Thm 95}
Fix $C=2$, and consider any rate $\mu$ and linear demand function $\lambda(\cdot)$. Let $\pi^*$ denote a revenue maximizing dynamic policy.  Then there exists a static pricing policy $\pi_s$ such that 
\begin{equation*}
\frac{\ObjP^{\pi^s}}{\ObjP^{\pi^*}}\geq 0.955.
\end{equation*}
\end{theorem}
This result establishes that for profit maximization, a simple static pricing policy guarantees more than 95.5\% of an optimal dynamic pricing policy. This is a much higher guarantee than for the general multi-objective case. In particular, for profit maximization, there is extremely limited value in dynamic pricing.

We note that due to the technical difficulty of the analysis, our result is limited to the case with only 2 units ($C=2$), and when the demand is linear $(p(\cdot)$ and $\lambda(\cdot)$ are linear), a common assumption in both the literature and practice. \edit{However, in Section \ref{NumEx}, we  see numerically that the level of guarantee above appears valid beyond the case $C=2$ and linear demand. \textit{In fact, our computational results illustrate that the 95.5\% lower bound holds across every single instance tested in a wide testbed  (across values of $C$ and demand models).}}

The proof of Theorem \ref{Thm 95} is again constructive in that it exhibits a particular static policy with such a guarantee. This policy is the same as the one presented in Eq \eqref{eq:lambda}. The proof relies on lower bounding the ratio of the service levels, which is indeed a lower bound on the profit ratio as seen in Eq. \eqref{eq:rR}. Then, the first order conditions of the profit maximization objective are used to impose constraints on the worst-case arrival rates of an optimal policy, which allows us to find a tighter bound on the ratio of the service levels.

\begin{proof}{Proof of Theorem \ref{Thm 95}.}
Let $\lambda_1^*, \lambda_2^*$ be the effective  arrival rates under the optimal policy for profit maximization, and $p_1^*, p_2^*$ be the corresponding optimal prices. Let $z_1=\frac{\lambda_1^* \lambda_2^*}{\mu^2}$ and $z_2=\frac{\lambda_2^*}{\mu}$. For the static policy, let $\tilde{\lambda}$ be defined according to \eqref{eq:lambda}. Since $C=2$, by Eq. \eqref{eq:rR} we have
\begin{equation*}
\frac{\ObjP^{\tilde{\lambda}}}{\ObjP^*} \geq\frac{1-\mathbb{P}_0(\tilde{\lambda})}{1-\mathbb{P}_0^*} = \frac{z_1^{2}+4z_1z_2+3z_1+4z_2^{2}+6z_2+2}{z_1^{2}+4z_1z_2+2z_1+5z_2^{2}+6z_2+2}:=R(z_1,z_2).
\end{equation*}

Next, we show that $R(z_1,z_2)$ is increasing in $z_1$ and decreasing in $z_2$ by simply looking at the first partial derivatives. Taking derivatives of $R$ w.r.t $z_1$ and $z_2$ gives
\begin{align*}
\frac{\partial R(z_1,z_2)}{\partial z_1} &= \frac{-z_1^2+2z_1z_2^2+4z_2^3+7z_2^2+6z_2+2}{(z_1^2+4z_1z_2+2z_1+5z_2^2+6z_2+2)^2} \geq 0\\
\frac{\partial R(z_1,z_2)}{\partial z_2} & = -\frac{2(z_1^2(z_2+2)+z_1(2z_2^2+7z_2+3)+z_2(3z_2+2))}{(z_1^2+4z_1z_2+2z_1+5z_2^2+6z_2+2)^2} \leq 0.
\end{align*}
To see that the partial derivative w.r.t. $z_1$ is  non-negative, it is sufficient to show that $z_1 \leq z_2^2$, which follows from the fact that  $\lambda_1^* \leq\lambda_2^*$, established in Lemma \ref{lemma Structural}. To see that the partial derivative w.r.t. $z_2$ is  non-positive, observe that all terms in the numerator are negative.

The remainder of the proof proceeds by dividing the analysis in two cases: if  $z_2$ is above or below $ \frac{\sqrt[]{7}-1}{3}$.  When $z_2\leq\frac{\sqrt[]{7}-1}{3}$,  then in this case
\begin{equation*}
R(z_1,z_2) \geq R(0,\frac{\sqrt[]{7}-1}{3}) \approx 0.9557
\end{equation*}
since  $R(z_1,z_2)$ is increasing in $z_1$ and decreasing in $z_2$.

For the remainder of the proof we consider the case where $z_2>\frac{\sqrt[]{7}-1}{3}$. In this case, we  leverage the first-order optimality conditions of the problem to show in Lemma \ref{LemOptimality} that $z_1$ and $z_2$ must be within a provable quantity of one another. This constraint then allows us to tighten the lower bound on $R(\cdot)$. Denote $\gamma_i=-p'(\lambda_i)$. Notice that since the demand is linear, then $\gamma_1=\gamma_2$. Define $\beta :=\frac{p_1^*-c}{\gamma_1}\geq 0$, and now we are ready to state the bounds on $z_1$ and $z_2$ in Lemma \ref{LemOptimality} (proved in Section \ref{Proofs}).

\begin{lemma}\label{LemOptimality}
Let $g(\beta,z_2)=\sqrt[]{(z_2+1)^2+\beta z_2(z_2+2)}-(z_2+1)$. Then 
\begin{align}
z_1&\geq g(\beta,z_2) \label{z1Opt}\\
z_2&\leq \sqrt[]{2\beta} \label{z2Opt}.
\end{align}
\end{lemma}

By Lemma \ref{LemOptimality} and the fact that $R(z_1,z_2)$ is non-decreasing in $z_1$ we have that
\begin{align}
R(z_1,z_2)&\geq R(g(\beta,z_2),z_2) \nonumber \\
&=\frac{(2+\beta)z_2^{2}+(2\beta+3)z_2+(2z_2+1)\sqrt[]{(1+\beta)z_2^{2}+2(1+\beta)z_2+1}+1}{(3+\beta)z_2^{2}+(2\beta+4)z_2+2z_2\sqrt[]{(1+\beta)z_2^{2}+2(1+\beta)z_2+1}+2} \nonumber \\
&=1 - \frac{z_2^{2}+z_2+1-\sqrt[]{(1+\beta)z_2^{2}+2(1+\beta)z_2+1}}{(3+\beta)z_2^{2}+(2\beta+4)z_2+2z_2\sqrt[]{(1+\beta)z_2^{2}+2(1+\beta)z_2+1}+2} \nonumber \\
&:=1-G(\beta,z_2). \label{eq:G}
\end{align}
Therefore, minimizing $R(z_1,z_2)$ is equivalent to maximizing $G(\beta,z_2)$, for which we provide an upper bound in Lemma \ref{Max G} (proved in Section \ref{Proofs}).

\begin{lemma}\label{Max G}
If $z_2\geq \frac{\sqrt[]{7}-1}{3}$ and $\beta \geq 0$, then $G(\beta,z_2)\leq 0.0433$.
\end{lemma}

Therefore, in the case when $z_2\geq \frac{\sqrt[]{7}-1}{3}$, Eq. \eqref{eq:G} and Lemma \ref{Max G} imply that
\begin{equation*}
R(z_1,z_2)\geq 1- G(\beta,z_2) \geq 1-0.0433 = 0.9567. 
\end{equation*} 
Combining both cases, we obtain the claimed result and  the proof is complete.  
\Halmos
\end{proof}

%% file: 7_Conclusion.tex
\section{Conclusion}\label{Conclusion}
\edit{
In this paper, we have provided the first universal guarantees on the strength of static pricing for reusable resources. Namely, we show that 78.9\% of the  profit, market share, and service level from the optimal dynamic pricing policy can be obtained by a static price. Our proof relies on a novel construction where the static arrival rate is set to the expected arrival rate of the optimal policy when there is at least one unit available. We sharpen the bound to 95.5\% in a special case where there are two units, demand is linear, and profit is being maximized. We believe that our static pricing policy construction naturally leads to analyzing a ratio of stock-in probabilities in various more general settings, although analyzing the ratio may require new ideas.} 

\edit{One important extension of this model is to allow for general i.i.d. service times rather than exponential service times.  In such a setting, the optimal dynamic pricing policy may no longer simply depend on the inventory state, as the remaining service time of every unit must also be tracked in the optimal dynamic pricing policy. However, if one restricts attention to inventory-based dynamic pricing policies, then our results have the potential to be generalized (see, e.g., \cite{brumelle1978generalization}, for an analysis of steady state probabilities of inventory state-dependent Erlang loss models). }

\edit{Other important extensions of the results naturally include ones with multiple classes of customers, products, and resource types. One may also consider the resources moving in a network to model ride-sharing applications in more detail. A model with  non-stationary arrivals may also be of interest, in which case the hope would be to prove that a price that only depends on time, but not the inventory state, is near-optimal.}

%% file: 6_Numeric.tex
\section{Numerical Experiments}\label{NumEx}
In this section, we conduct a set of numerical experiments to test the performance of the static pricing policy.  We consider three types of demand functions:  linear, exponential, and logistic. For a linear demand curve, we assume it takes the form $\lambda=-ap+b$; for an exponential demand curve, we assume it follows $\lambda=be^{-ap}$; for the logistic demand curve, we assume it is $\lambda=\frac{b(1+e^{-ap^0})}{1+e^{a(p-p^0)}}$ where $p^0$ is the inflection point. Notice that in all three demand curves, the maximum demand rate is set to be $b$ when the price is set to 0.

For each value of $C$, we randomly generate the mean usage time uniformly in $\frac{1}{\mu}\in [0.05,50]$; $a$ is randomly generated uniformly between 0.1 and 5; $b$ is randomly generated uniformly between 0.5 and 10; $p^0$ is randomly generated uniformly between [0,20]. We assume that the average service cost is 0, i.e., $c=0$. We generate 1,000 different instances of inputs and calculate the profit rate under the optimal dynamic pricing policy, the constructed static price policy $\tilde{\lambda}$ according to Eq. \eqref{eq:lambda}, and the best static price policy $\pi^{s^*}$. We report the \textit{worst case} of $\frac{\ObjP^{\tilde{\lambda}}}{\ObjP^{\pi^*}}$ and $\frac{\ObjP^{\pi^{s^*}}}{\ObjP^{\pi^*}}$ for each capacity level $C$. The results are summarized in  Table \ref{Table profit only}.

\begin{table}[h!]
\centering
\begin{tabular}{c|cc|cc|cc}
  \multicolumn{1}{c}{} & \multicolumn{2}{c}{Linear} & \multicolumn{2}{c}{Exponential} & \multicolumn{2}{c}{Logistic} \\
   \hline
$C$  & $\frac{\ObjP^{\tilde{\lambda}}}{\ObjP^{\pi^*}}$      & $\frac{\ObjP^{\pi^{s^*}}}{\ObjP^{\pi^*}}$       & $\frac{\ObjP^{\tilde{\lambda}}}{\ObjP^{\pi^*}}$        & $\frac{\ObjP^{\pi^{s^*}}}{\ObjP^{\pi^*}}$          & $\frac{\ObjP^{\tilde{\lambda}}}{\ObjP^{\pi^*}}$       & $\frac{\ObjP^{\pi^{s^*}}}{\ObjP^{\pi^*}}$        \\
\hline
2  & 99.53\% & 99.54\% & 99.06\% & 99.07\% & 99.16\% & 99.18\% \\
3  & 99.27\% & 99.28\% & 98.57\% & 98.60\% & 98.68\% & 98.72\% \\
4  & 99.10\% & 99.12\% & 98.26\% & 98.31\% & 98.41\% & 98.46\% \\
5  & 98.97\% & 99.00\% & 98.05\% & 98.11\% & 98.19\% & 98.28\% \\
10 & 98.66\% & 98.71\% & 97.58\% & 97.70\% & 97.71\% & 97.84\% \\
20 & 98.46\% & 98.55\% & 97.38\% & 97.56\% & 97.46\% & 97.70\% \\
30 & 98.40\% & 98.51\% & 97.39\% & 97.57\% & 97.45\% & 97.68\% \\
40 & 98.38\% & 98.51\% & 97.48\% & 97.62\% & 97.51\% & 97.72\% \\
50 & 98.37\% & 98.51\% & 97.60\% & 97.69\% & 97.58\% & 97.79\%
\end{tabular}
\caption{Worst case profit ratio: static pricing policies vs. optimal dynamic pricing policy.}
\label{Table profit only}
\end{table}
As one can observe, the performance of static pricing is generally higher than 97.5\%. When $C=2$, we observe the worst case to be 99.53\% in the case of linear demand, which is even higher than the 95.5\% guarantee proven in Theorem \ref{Thm 95}. We also note that this very high performance of static prices  continues to hold when we depart from the exact assumptions of Theorem \ref{Thm 95},  for general values of $C$ and for  exponential and logistic demand curves.

In general, the worst case performance of static pricing (either the best static price or the price we construct) does not happen when $C=2$. However, the ratio of the profit rate achieved by the static policy and the optimal profit rate appears to be relatively independent of the value of $C$. Of course, as $C$ approaches infinity, the worst case ratio indeed converges to 1.

In addition, one may observe that the performance of the static price policy we constructed in the proofs is very close to the performance of the best static price for profit maximization. The difference of the worst case performance between the two static prices is usually less than 0.2\%.

Using a similar testbed, we also conducted numerical experiments for the multi-objective case. We use the linear demand model in the numerical experiment and randomly generate the values of $\alpha_i$'s uniformly at random. The rest of the experiment settings are the same as described before. We calculate the worst case performance of our constructed static price compared to the total objective as well as  for the three performance metrics. The results are presented in Table \ref{Table multi}.

\begin{table}[h!]
\centering
\begin{tabular}{c|c|c|c|c}
C  & $\frac{V^{\tilde{\lambda}}}{V^*}$ & $\frac{\ObjP^{\tilde{\lambda}}}{\ObjP^{\pi^*}}$ & $\frac{\ObjM^{\tilde{\lambda}}}{\ObjM^{\pi^*}}$ & $\frac{\ObjA^{\tilde{\lambda}}}{\ObjA^{\pi^*}}$ \\
\hline
2  & 81.08\% & 84.70\% & 81.03\% & 81.03\% \\
3  & 80.32\% & 83.85\% & 80.23\% & 80.23\% \\
4  & 80.95\% & 84.45\% & 80.82\% & 80.82\% \\
5  & 81.80\% & 85.27\% & 81.63\% & 81.63\% \\
10 & 85.37\% & 88.67\% & 85.02\% & 85.02\% \\
15 & 87.68\% & 90.79\% & 87.17\% & 87.17\% \\
20 & 89.30\% & 92.22\% & 88.67\% & 88.67\%
\end{tabular}
\caption{Performance of static pricing with multiple objectives.}
\label{Table multi}
\end{table}
As one may notice, the lowest of the worst case performance ratio happens when $C=3$ at 80.32\% for the  overall objectives, and 80.23\% for the market share and service level objectives. For this worst case ratio, the values of $\alpha_i$'s are similar to the construction in our tightness example where $\alpha_3$ is very close to 1 while $\alpha_1$ and $\alpha_2$ is close to 0. This finding is consistent with our tightness analysis.

%% file: 8_Appendix.tex
\section{Additional proofs}\label{Proofs}

\begin{proof}{\underline{\textbf{Proof of Lemma \ref{lemma Structural}}}.}
We prove this lemma by transforming the continuous time Markov Decision Process (MDP) to a discrete time MDP and showing that the value iteration operator preserve concavity and monotonicity.

Using standard techniques (see, e.g.,  \citet{bertsekas2005dynamic2}), the continuous time MDP associated to Equation (\ref{Eq1}) can be transformed into a discrete time MDP, through uniformization,  and solved efficiently using value iteration. Let $\gamma$ be given by
\begin{equation*}
   \gamma = \frac{1}{1+\Lambda+C\mu},
\end{equation*}
where $\Lambda$ is the maximum demand rate. Note that $1+\Lambda+C\mu$ upper bounds the transition rates from any state in the Markov Chain. 

Let $h(i)$ denote the relative, long-run expected reward associated with having $i$ units available and $\eta$ be the optimal average profit. The value iteration operator, $\mathcal{T}$, takes the following form,
\begin{multline}\label{Eq2}
\mathcal{T}h(i) = \max_{\lambda \in [0,\Lambda]} \{\alpha_1\lambda(p(\lambda)-c)+\alpha_2\lambda+\alpha_3-\eta+\\ \gamma \lambda h(i-1)+\gamma \mu (C-i)h(i+1) + (1-\gamma(\lambda+\mu(C-i))h(i)) \}\ \forall i
\end{multline}
where
\begin{align*}
&h(0) = 0.
\end{align*}

Letting $h^*(i)$ denote the relative optimal expected reward of having $i$ units available, then $h^\ast(i) = \lim\limits_{n \to \infty}\mathcal{T}^nh(i)$. We next prove the fact that $h^*(i)$ is nondecreasing and concave by showing $\mathcal{T}h(i)$ is nondecreasing and concave if $h(i)$ has the same properties.  

 For any state $i$, we can rewrite the value iteration presented in Equation (\ref{Eq2}) as follows:
\begin{equation*}
\mathcal{T}h(i) = A(i)+B(i)
\end{equation*}
where
\begin{align*}
A(i) &= \max_{\lambda \in [0,\Lambda]} \left[\alpha_1\lambda(p(\lambda)-c)+\alpha_2\lambda+\gamma \lambda h(i-1)+ \gamma(1+\Lambda-\lambda)h(i)\right],\\
B(i) &= \gamma \mu \left[(C-i)h(i+1)+i h(i)\right]-\eta+\alpha_3 .
\end{align*}
Denote 
\begin{equation*}
\lambda_i = \arg\max A(i)
\end{equation*}
In order to show $\mathcal{T}h(i)$ is nondecreasing and concave, we will show both $A(i)$ and $B(i)$ are nondecreasing and concave.

To show that $A(i)$ is nondecreasing in $i$, observe that     
\begin{align*}
A(i)-A(i-1) &= A(i)|_{\lambda_{i}} - A(i-1)|_{\lambda_{i-1}}\\
& \geq A(i)|_{\lambda_{i-1}} - A(i-1)|_{\lambda_{i-1}}\\
&=\gamma\lambda_{i-1}\left[h(i-1)-h(i-2)\right] + \gamma (1+\Lambda-\lambda_{i-1})\left[h(i)-h(i-1)\right]\\
&\geq 0.
\end{align*}
The first inequality comes from the fact that $\lambda_i$ is the maximizer of $A(i)$. The last inequality comes from the assumption that $h(\cdot)$ is nondecreasing.

To show that $B(i)$ in nondecreasing in $i$, observe that
	\begin{align*}
	B(i)-B(i-1) &=\gamma \mu \left( (C-i)h(i+1)+i h(i)-(C-i+1)h(i)-(i-1)h(i-1)\right)\\
	&= \gamma \mu \left((C-i)\left[h(i+1)-h(i)\right] + (i-1)\left[h(i)-h(i-1)\right]\right)\\
	&\geq 0,
	\end{align*}
	since $h(\cdot)$ is nondecreasing and both $\gamma$ and $\mu$ are positive.

To establish the concavity of $A(i)$, observe that
    \begin{align*}
	&A(i-1)+A(i+1)-2A(i)\\ 
	&= A(i-1)|_{\lambda_{i-1}} + A(i+1)|_{\lambda_{i+1}} - 2A(i)|_{\lambda_{i}}\\
	&\leq A(i-1)|_{\lambda_{i-1}} + A(i+1)|_{\lambda_{i+1}} - A(i)|_{\lambda_{i-1}} - A(i)|_{\lambda_{i+1}}\\
	&= \gamma \lambda(\lambda_{i-1})h(i-2) + \gamma(1+\Lambda-\lambda_{i-1})h(i-1)
	+\lambda_{i+1}h(i) + \gamma(1+\Lambda-\lambda_{i+1})h(i+1)\\
	&\ \ -\lambda_{i-1}h(i-1) + \gamma(1+\Lambda-\lambda_{i-1})h(i)
	-\lambda_{i+1}h(i-1) + \gamma(1+\Lambda-\lambda_{i+1})h(i)\\
	&=\gamma \lambda_{i-1}\left[h(i-2)+h(i)-2h(i-1)\right]
	+\gamma(1+\Lambda)\left[h(i-1)+h(i+1)-2h(i)\right]\\
	&\ \ + \gamma \lambda_{i+1}\left[2h(i)-h(i-1)-h(i+1)\right]
	\\
	&= \gamma \lambda_{i-1}\left[h(i-2)+h(i)-2h(i-1)\right] + \gamma (1+\Lambda-\lambda_{i+1})\left[h(i-1)+h(i+1)-2h(i)\right]\\
	&\leq 0.
	\end{align*}
	The first inequality follows from the fact that $\lambda_i$ is the maximizer of $A(i)$. Since $\Lambda$ is the maximum rate of customer arrivals, then $1+\Lambda-\lambda_{i+1}$ is positive and the last inequality follows from the concavity of $h(\cdot)$.

 To establish the concavity of $B(i)$, observe that
	\begin{align*}
	B(i-1)+B(i+1)-2B(i)
	&=\gamma \mu \left[(C-(i-1))h(i)+(i-1)h(i-1)\right]\\
	&\ \  +\gamma \mu \left[(C-(i+1))h(i+2) + (i+1)h(i+1)\right]\\
	&\ \  - \gamma \mu \left[2(C-i)h(i+1)-2i h(i)\right]
	\\
	&=\gamma \mu \left[(i-1)\left[h(i-1)+h(i+1)-2h(i)\right]\right]\\&\ \  +\gamma \mu \left[(C-i-1)\left[h(i+2)+h(i)-2h(i+1)\right]\right]\\
	&\leq 0.
	\end{align*}
	The last inequality follows from the assumption that $h(\cdot)$ is concave and the fact that both $\gamma$ and $\mu$ are positive. 
Recall from Equation (\ref{Eq2}), the optimal prices can be solved using the following equation,
\begin{equation*}
\lambda_i^* = \arg\max_\lambda \ \lambda[\alpha_1(p(\lambda)-c)+\alpha_2-\gamma(h^*(i)-h^*(i-1)].
\end{equation*}
Given the nondecreasing and concave properties of $h^*(\cdot)$, we can conclude the desired property of the optimal policy.
\Halmos
\end{proof}

\begin{proof}{\underline{\textbf{Proof of Lemma \ref{Derivative}}}.}
The proof follows by simply showing that $\frac{\partial \tilde{R}(z_1,\ldots,z_C)}{\partial z_1}\geq 0$, which is equivalent to showing that the numerator of   $\frac{\partial \tilde{R}(z_1,\ldots,z_C)}{\partial z_1}$ is non-negative. To do this, we first establish a few facts.  Recall that $a_i := \frac{C!}{(C-i)!}$ for $i=0,1,\ldots, C$ and $z_i:= \Pi_{j=i}^C \frac{\lambda_j^*}{\mu}$ for $i=1, \ldots, C+1$. Also recall that $x := \sum_{k=1}^{C} a_kz_{k+1}$ and $y := \sum_{k=2}^{C} a_k z_k$.

Since $\lambda_i^*$ is non-decreasing in $i$ from Lemma \ref{lemma Structural}, then for $k=1,\ldots, C$ we have that
\begin{align}
z_1z_{k+1} = \Pi_{i=1}^C\frac{\lambda^*_i}{\mu} \Pi_{j=k+1}^C\frac{\lambda^*_j}{\mu}
\leq \Pi_{i=2}^C\frac{\lambda^*_i}{\mu} \Pi_{j=k}^C \frac{\lambda^*h_j}{\mu}
=z_2z_k. \label{eq:3.1}
\end{align}
Therefore,
\begin{align}
y(a_1z_1+y) =\left(\sum_{j=2}^C a_jz_j\right) \left(\sum_{i=1}^C a_iz_i\right)
\geq a_2z_2\left(\sum_{i=1}^C a_iz_i\right)
\geq a_2\left(\sum_{i=1}^C a_iz_1z_{i+1}\right)
=a_2z_1x. \label{eq:3.2}
\end{align}
where the second inequality follows from Eq. \eqref{eq:3.1}.

Under the assumption of $y\geq a_1z_1$ and the fact that $y\leq (C-1)x$, we also have that 
\begin{align}
x&\geq z_1  \label{eq:3.3},\\
x&\geq\frac{a_1z_1+y}{2(C-1)}. \label{eq:3.4}
\end{align}
Using the definitions of $x$ and $y$, we may rewrite $\tilde{R}(\cdot)$ as 
\begin{align*}
\tilde{R}(z_1,\ldots,z_C)&=\frac{(a_0z_1+x)[a_1(a_1z_1+y)^3+a_2(a_1z_1+y)^2x+a_3(a_1z_1+y)x^2+a_4x^3]}{a_0(a_1z_1+y)^4+a_1(a_1z_1+y)^3x+a_2(a_1z_1+y)^2x^2+a_3(a_1z_1+y)x^3+a_4x^4}.
\end{align*}
The derivative of the numerator of $\tilde{R}(z_1,\ldots,z_C)$is
\small
\begin{align*}
&[(a_1z_1+y)^4+a_1(a_1z_1+y)^3x+a_2(a_1z_1+y)^2x^2+a_3(a_1z_1+y)x^3+a_4x^4]\times\\
&[3a_1^2(a_1z_1+y)^2(z_1+x)+2a_1a_2(a_1z_1+y)(z_1+x)x+a_1a_3(z_1+x)x^2\\&\ +a_1(a_1z_1+y)^3+a_2(a_1z_1+y)^2x+a_3(a_1z_1+y)x^2+a_4x^3]\\
&-[a_1(a_1z_1+y)^3(z_1+x)+a_2(a_1z_1+y)^2(z_1+x)x+a_3(a_1z_1+y)(z_1+x)x^2+a_4(z_1+x)x^3]\times\\
&[4a_1(a_1z_1+y)^3+3a_1^2(a_1z_1+y)^2x+2a_1a_2(a_1z_1+y)x^2+a_1a_3x^3]\\
\\
=&3a_1^2(a_1z_1+y)^6(z_1+x)+2a_1a_2(a_1z_1+y)^5(z_1+x)x+a_1a_3(a_1z_1+y)^4(z_1+x)x^2\\
&\ +a_1(a_1z_1+y)^7+a_2(a_1z_1+y)^6x+a_3(a_1z_1+y)^5x^2+a_4(a_1z_1+y)^4x^3\\
&+3a_1^3(a_1z_1+y)^5(z_1+x)x+2a_1^2a_2(a_1z_1+y)^4(z_1+x)x^2+a_1^2a_3(a_1z_1+y)^3(z_1+x)x^3\\
&\ +a_1^2(a_1z_1+y)^6x+a_1a_2(a_1z_1+y)^5x^2+a_1a_3(a_1z_1+y)^4x^3+a_1a_4(a_1z_1+y)^3x^4\\
&+3a_1^2a_2(a_1z_1+y)^4(z_1+x)x^2+2a_1a_2^2(a_1z_1+y)^3(z_1+x)x^3+a_1a_2a_3(a_1z_1+y)^2(z_1+x)x^4\\
&\ +a_1a_2(a_1z_1+y)^5x^2+a_2^2(a_1z_1+y)^4x^3+a_2a_3(a_1z_1+y)^3x^4+a_2a_4(a_1z_1+y)^2x^5\\
&+3a_1^2a_3(a_1z_1+y)^3(z_1+x)x^3+2a_1a_2a_3(a_1z_1+y)^2(z_1+x)x^4+a_1a_3^2(a_1z_1+y)(z_1+x)x^5\\
&\ +a_1a_3(a_1z_1+y)^4x^3+a_2a_3(a_1z_1+y)^3x^4+a_3^2(a_1z_1+y)^2x^5+a_3a_4(a_1z_1+y)x^6\\
&+3a_1^2a_4(a_1z_1+y)^2(z_1+x)x^4+2a_1a_2a_4(a_1z_1+y)(z_1+x)x^5+a_1a_3a_4(z_1+x)x^6\\
&\ +a_1a_4(a_1z_1+y)^3x^4+a_2a_4(a_1z_1+y)^2x^5+a_3a_4(a_1z_1+y)x^6+a_4^2x^7\\
&-4a_1^2(a_1z_1+y)^6(z_1+x)-3a_1^3(a_1z_1+y)^5(z_1+x)x-2a_1^2a_2(a_1z_1+y)^4(z_1+x)x^2-a_1^2a_3(a_1z_1+y)^3(z_1+x)x^3\\
&-4a_1a_2(a_1z_1+y)^5(z_1+x)x-3a_1^2a_2(a_1z_1+y)^4(z_1+x)x^2-2a_1a_2^2(a_1z_1+y)^3(z_1+x)x^3-a_1a_2a_3(a_1z_1+y)^2(z_1+x)x^4\\
&-4a_1a_3(a_1z_1+y)^4(z_1+x)x^2-3a_1^2a_3(a_1z_1+y)^3(z_1+x)x^3-2a_1a_2a_3(a_1z_1+y)^2(z_1+x)x^4-a_1a_3^2(a_1z_1+y)(z_1+x)x^5\\
&-4a_1a_4(a_1z_1+y)^3(z_1+x)x^3-3a_1^2a_4(a_1z_1+y)^2(z_1+x)x^4-2a_1a_2a_4(a_1z_1+y)(z_1+x)x^5-a_1a_3a_4(z_1+x)x^6\\
\\
=&3a_1^2(a_1z_1+y)^6(z_1+x)+[2a_1a_2+3a_1^3](a_1z_1+y)^5(z_1+x)x+[a_1a_3+5a_1^2a_2](a_1z_1+y)^4(z_1+x)x^2+a_1(a_1z_1+y)^7\\
&+[a_2+a_1^2](a_1z_1+y)^6x+[a_3+2a_1a_2](a_1z_1+y)^5x^2+[a_4+a_2^2+2a_1a_3](a_1z_1+y)^4x^3+[4a_1^2a_3+2a_1a_2^2](a_1z_1+y)^3(z_1+x)x^3\\
&+[2a_1a_4+2a_2a_3](a_1z_1+y)^3x^4+[3a_1a_2a_3+3a_1^2a_4](a_1z_1+y)^2(z_1+x)x^4+[2a_2a_4+a_3^2](a_1z_1+y)^2x^5\\
&+[a_1a_3^2+2a_1a_2a_4](a_1z_1+y)(z_1+x)x^5+2a_3a_4(a_1z_1+y)x^6+a_1a_3a_4(z_1+x)x^6+a_4^2x^7\\
&-4a_1^2(a_1z_1+y)^6(z_1+x)-[3a_1^3+4a_1a_2](a_1z_1+y)^5(z_1+x)x-[5a_1^2a_2+4a_1a_3](a_1z_1+y)^4(z_1+x)x^2\\
&-[4a_1^2a_3+2a_1a_2^2+4a_1a_4](a_1z_1+y)^3(z_1+x)x^3-[3a_1a_2a_3+3a_1^2a_4](a_1z_1+y)^2(z_1+x)x^4\\
&-[a_1a_3^2+2a_1a_2a_4](a_1z_1+y)(z_1+x)x^5-a_1a_3a_4(z_1+x)x^6\\
\\
=&a_1(a_1z_1+y)^7+[a_2+a_1^2](a_1z_1+y)^6x+[a_3+2a_1a_2](a_1z_1+y)^5x^2+[a_4+a_2^2+2a_1a_3](a_1z_1+y)^4x^3\\
&+[2a_1a_4+2a_2a_3](a_1z_1+y)^3x^4+[2a_2a_4+a_3^2](a_1z_1+y)^2x^5+2a_3a_4(a_1z_1+y)x^6+a_4^2x^7\\
&-a_1^2(a_1z_1+y)^6(z_1+x)-2a_1a_2(a_1z_1+y)^5(z_1+x)x-3a_1a_3(a_1z_1+y)^4(z_1+x)x^2-4a_1a_4(a_1z_1+y)^3(z_1+x)x^3\\
\\
=&a_1(a_1z_1+y)^7+a_2(a_1z_1+y)^6x+a_3(a_1z_1+y)^5x^2+[a_4+a_2^2-a_1a_3](a_1z_1+y)^4x^3\\
&+[2a_2a_3-2a_1a_4](a_1z_1+y)^3x^4+[2a_2a_4+a_3^2](a_1z_1+y)^2x^5+2a_3a_4(a_1z_1+y)x^6+a_4^2x^7\\
&-a_1^2(a_1z_1+y)^6z_1-2a_1a_2(a_1z_1+y)^5z_1x-3a_1a_3(a_1z_1+y)^4z_1x^2-4a_1a_4(a_1z_1+y)^3z_1x^3\\
\\
=&a_1y(a_1z_1+y)^6+a_2y(a_1z_1+y)^5x+a_3y(a_1z_1+y)^4x^2+[a_4+a_2^2-a_1a_3]y(a_1z_1+y)^3x^3\\
&+[2a_2a_3-2a_1a_4](a_1z_1+y)^3x^4+[2a_2a_4+a_3^2](a_1z_1+y)^2x^5+2a_3a_4(a_1z_1+y)x^6+a_4^2x^7\\
&-a_1a_2(a_1z_1+y)^5z_1x-2a_1a_3(a_1z_1+y)^4z_1x^2-[3a_1a_4-a_1a_2^2+a_1^2a_3](a_1z_1+y)^3z_1x^3\\
\\
\geq&a_1a_2(a_1z_1+y)^5z_1x+a_2^2(a_1z_1+y)^4z_1x^2+a_1a_3(a_1z_1+y)^4z_1x^2+[a_1a_4+a_1a_2^2-a_1^2a_3](a_1z_1+y)^3z_1x^3\\
&+[2a_2a_3-2a_1a_4](a_1z_1+y)^3z_1x^3+\frac{2a_2a_4+a_3^2}{2(C-1)}(a_1z_1+y)^3z_1x^3+\frac{a_3a_4}{2(C-1)^2}(a_1z_1+y)^3z_1x^3+a_4^2x^7\\
&-a_1a_2(a_1z_1+y)^5z_1x-2a_1a_3(a_1z_1+y)^4z_1x^2-[3a_1a_4-a_1a_2^2+a_1^2a_3](a_1z_1+y)^3z_1x^3\\
\\
\geq&0
\end{align*}
\normalsize
The first equality follows from expanding the products completely. The second equality follows from combining positive terms, and then the negative terms. The third equality follows from canceling terms out. The fourth equality follows from expanding $(z_1+x)$ terms and simplifying. The fifth equality follows form expanding $(a_1z_1+y)$ in some of the positive terms and simplifying. The first inequality follows from lower bounding some terms using $y\geq a_1z_1$, Eq. \eqref{eq:3.2}, Eq. \eqref{eq:3.3}, or Eq. \eqref{eq:3.4}. The second inequality follows since 
\begin{align*}
a_2^2=C^2(C-1)^2\geq C^2(C-1)(C-2)= a_1a_3
\end{align*}
and
\begin{align*}
&a_1a_4+a_1a_2^2-a_1^2a_3+2a_2a_3-2a_1a_4+\frac{2a_2a_4+a_3^2}{2(C-1)}+\frac{a_3a_4}{2(C-1)^2}-[3a_1a_4-a_1a_2^2+a_1^2a_3]\\
=&2a_1a_2^2-2a_1^2a_3-4a_1a_4+2a_2a_3+\frac{2a_2a_4+a_3^2}{2(C-1)}+\frac{a_3a_4}{2(C-1)^2}\\
=&C^2(6+C(6C-13))\\
>&0 
\end{align*}
when $C\geq 4$.
\Halmos
\end{proof}

\begin{proof}{\underline{\textbf{Proof of Lemma \ref{Cal bound}}}.}
Recall that\begin{equation*}
\tilde{R}(0,z_2,\ldots,z_C) = \frac{\sum_{i=1}^4 a_i y^{4-i}x^{i}}{\sum_{i=0}^4 a_i y^{4-i}x^i}
\end{equation*}

The main idea in proving this lemma is to compare the ratio of the coefficient of every term in $\tilde{R}(0,z_2,\ldots,z_C)$. First, we restrict our focus only to the ratio of the coefficients for the terms when $y^4$ is expanded, since the ratio of the coefficients terms not in $y^4$ is 1. To see the fact that every term not in $y^4$ has the same value of coefficient in both the numerator and denominator, we can rewrite $\tilde{R}(0,z_2,\ldots,z_C)$ as 
\begin{equation*}
\tilde{R}(0,z_2,\ldots,z_C) =\frac{\sum_{i=1}^4 a_i y^{4-i}x^{i}}{y^4+\sum_{i=1}^4 a_i y^{4-i}x^{i}}
\end{equation*}
Since every term  not in $y^4$ must be in $\sum_{i=1}^4 a_i y^{4-i}x^{i}$, and $\sum_{i=1}^4 a_i y^{4-i}x^{i}$ appears in both numerator and denominator, then the ratio of the coefficient of the terms not in $y^4$ must be 1. Since $\tilde{R}(0,z_2,\ldots,z_C)\leq 1$ by definition, we only need to focus on the ratio of the coefficient of the terms in $y^4$ to find the lower bound of $\tilde{R}(0,z_2,\ldots,z_C)$.

We now calculate a lower bound on the ratio of the coefficients for the  $y^4$ terms. By the definition of $y=\sum_{k=2}^C a_kz_k$, every term in $y^4$ takes the form: $z_2^{k_2}\cdots z_C^{k_C}$ where $\sum_{i=2}^C k_i=4, k_i\in \mathbb{N}$. Therefore, a combination $(k_2,\ldots,k_C)$ uniquely defines a term in $y^4$. For $i=1,\ldots,4$, we use the set $\mathcal{S}_i$ to select possible ways of choosing terms from $y$ and $x$, and is defined as
\begin{align*}
\mathcal{S}_i = \{k',k'':&\sum_{j=2}^C k_j'=4-i,\\
&\sum_{j=2}^C k_j''=i,\\
&k_j'+k_j''=k_j,\ j=2,\ldots,C\\
&k_j',k_j''\in \mathbb{N}\}.
\end{align*}

Now let $A(k_2,\ldots,k_C)$ denote the coefficient of the term defined by $(k_2,\ldots,k_C)$ in the numerator and $B(k_2,\ldots,k_C)$ denote the coefficient of that term in the denominator. Plugging in $y=\sum_{k=2}^C a_kz_k$ and $x=\sum_{k=1}^C a_kz_{k+1}$ into $\tilde{R}(0,z_2,\ldots,z_C)$, we have that
\begin{align*}
A(k_2,\ldots,k_C)&=\sum_{i=1}^{4}a_i\left[\sum_{k',k''\in \mathcal{S}_i}\frac{(4-i)!}{k_2'!\cdots k_C'!}\Pi_{j=2}^{C}\left(\frac{C!}{(C-j)!}\right)^{k_j'}\frac{i!}{k_2''!\cdots k_C''!}\Pi_{j=2}^{C}\left(\frac{C!}{(C-j+1)!}\right)^{k_j''}\right]\\
B(k_2,\ldots,k_C)&=\frac{4!}{k_2!\cdots k_C!}\Pi_{j=2}^{C}\left(\frac{C!}{(C-j)!}\right)^{k_j}+A(k_2,\ldots,k_C).
\end{align*}
Notice that $A(k_2,\ldots,k_C)$ is from $\sum_{i=1}^4 a_iy^{4-i}x^i$, and $\frac{4!}{k_2!\cdots k_C!}\Pi_{j=2}^{C}\left(\frac{C!}{(C-j)!}\right)^{k_j}$ is from $y^4$.

Therefore, 
\begin{align*}
\tilde{R}(z_2,\ldots,z_C) &\geq \min \frac{A(k_2,\ldots,k_C)}{B(k_2,\ldots,k_C)}\\
&= \min \frac{A(k_2,\ldots,k_C)}{\frac{4!}{k_2!\cdots k_C!}\Pi_{j=2}^{C}\left(\frac{C!}{(C-j)!}\right)^{k_j}+A(k_2,\ldots,k_C)}\\
&= \min \frac{1}{\frac{\frac{4!}{k_2!\cdots k_C!}\Pi_{j=2}^{C}\left(\frac{C!}{(C-j)!}\right)^{k_j}}{A(k_2,\ldots,k_C)}+1}\\
&= \min \frac{1}{F(k_2,\ldots,k_C)+1}
\end{align*}
where
\begin{align*}
F(k_2,\ldots,k_C) = \frac{\frac{4!}{k_2!\cdots k_C!}\Pi_{j=2}^{C}\left(\frac{C!}{(C-j)!}\right)^{k_j}}{A(k_2,\ldots,k_C)}.
\end{align*}
To find the minimum of $\frac{A(k_2,\ldots,k_N)}{B(k_2,\ldots,k_N)}$ is equivalent to finding the maximum of $F(k_2,\ldots,k_C)$.

We  next show that for $C\geq 4$, $F(k_2,\ldots,k_C)$ is upper bounded by $\frac{27}{104}$. First, we show $F(k_2,\ldots,k_C)$ is maximized when $k_2=4$ and $k_i=0, \forall i=3,\ldots,C$. This corresponds to the term $z_2^4$. To see this, observe that
\begin{align*}
F(k_2,\ldots,k_C) &= \frac{\frac{4!}{k_2!\cdots k_C!}\Pi_{j=2}^{C}\left(\frac{C!}{(C-j)!}\right)^{k_j}}{\sum_{i=1}^{4}a_i\left[\sum_{k',k''\in \mathcal{S}_i}\frac{(4-i)!}{k_2'!\cdots k_C'!}\Pi_{j=2}^{C}\left(\frac{C!}{(C-j)!}\right)^{k_j'}\frac{i!}{k_2''!\cdots k_C''!}\Pi_{j=2}^{C}\left(\frac{C!}{(C-j+1)!}\right)^{k_j''}\right]}\\
&\leq\frac{\frac{4!}{k_2!\cdots k_C!}}{\sum_{i=1}^{4}\frac{C!}{(C-i)!}\frac{1}{(C-1)^i}\sum_{k',k''\in \mathcal{S}_i}\frac{(4-i)!}{k_2'!\cdots k_C'!}\frac{i!}{k_2''!\cdots k_C''!}}\\
&=\frac{\frac{4!}{k_2!\cdots k_C!}(C-1)^4}{\sum_{i=1}^{4}\frac{C!}{(C-i)!}(C-1)^{4-i}\sum_{k',k''\in \mathcal{S}_i}\frac{(4-i)!}{k_2'!\cdots k_C'!}\frac{i!}{k_2''!\cdots k_C''!}}\\
&=\frac{(C-1)^4}{\sum_{i=1}^{4}\frac{C!}{(C-i)!}(C-1)^{4-i}\frac{\sum_{k',k''\in \mathcal{S}_i}\frac{(4-i)!}{k_2'!\cdots k_C'!}\frac{i!}{k_2''!\cdots k_C''!}}{\frac{4!}{k_2!\cdots k_C!}}}\\
&=\frac{(C-1)^4}{\sum_{i=1}^{4}\frac{C!}{(C-i)!}(C-1)^{4-i}} \\
&:=H(C).\\
\end{align*}
The first equality is by the definition of $F(k_2,\ldots,k_C)$. The first inequality holds since for any $i$, the maximum possible ratio of the product terms in the numerator and the denominator is $(C-1)^i$. The second equality follows by multiplying the numerator and denominator by $(C-1)^4$. The third equality follows by dividing the numerator and denominator by $\frac{4!}{k_2!\cdots k_C!}$. The last equality holds because $\sum_{k',k''\in \mathcal{S}_i}\frac{(4-i)!}{k_2'!\cdots k_C'!}\frac{i!}{k_2''!\cdots k_C''!}$ and $\frac{4!}{k_2!\cdots k_C!}$ equivalent calculations of the same multinomial coefficient.

Next, we show that $H(C)$ is decreasing in $C$ for $C\geq 4$. Notice that
\begin{align*}
H(C+1)-H(C)&=\frac{C^4}{\sum_{i=4}^4 \frac{(C+1)!}{(C+1-i)!}C^{4-i}}-\frac{(C-1)^4}{\sum_{i=1}^{4}\frac{C!}{(C-i)!}(C-1)^{4-i}}\\
&= \frac{C^4\left(\sum_{i=1}^{4}\frac{C!}{(C-i)!}(C-1)^{4-i}\right)-(C-1)^4\left(\sum_{i=4}^4 \frac{(C+1)!}{(C+1-i)!}C^{4-i}\right)}{\left(\sum_{i=4}^4 \frac{(C+1)!}{(C+1-i)!}C^{4-i}\right)\left(\sum_{i=1}^{4}\frac{C!}{(C-i)!}(C-1)^{4-i}\right)}\\
&=\frac{-2C(C-1)[2C(C-1)(C-2)-1]}{\left(\sum_{i=4}^4 \frac{(C+1)!}{(C+1-i)!}C^{4-i}\right)\left(\sum_{i=1}^{4}\frac{C!}{(C-i)!}(C-1)^{4-i}\right)}\\
&\leq 0 \text{ for } C\geq 4.
\end{align*}
So, 
\begin{equation*}
F(k_2,\ldots,k_N)\leq H(C)\leq H(4) = \frac{27}{104}.
\end{equation*}

Therefore, we have the lower bound of $\tilde{R}(0,z_2,\ldots,z_C)$ as 
\begin{align*}
\tilde{R}(0,z_2,\ldots,z_C)&\geq \min \frac{T(k_2,\ldots,k_C)}{B(k_2,\ldots,k_C)}\\
&= \frac{1}{\max F(k_2,\ldots,k_C)+1}\\
&\geq \frac{1}{H(4)+1}\\
&= \frac{1}{\frac{27}{104}+1}\\
&=\frac{104}{131}.
\end{align*}
\Halmos
\end{proof}

\begin{proof}{\underline{\textbf{Proof of Lemma \ref{Cal bound2}}}.}

We directly calculate the lower bound of $\tilde{R}(z_1,\ldots,z_C)$. In this case, one can show that $z_C\geq C-1$ and $z_1\geq (C-1)^{C-1}$, so the $z_1^C$ dominates the rest of terms in $\tilde{R}(z_1,\ldots,z_C)$. Since the coefficient of $z_1^C$ is the same in the numerator and denominator, one can expect, in this case, $\tilde{R}(z_1,\ldots,z_C)$ to be close to 1.

First note that we have
\begin{equation*}
y = \sum_{k=2}^C a_kz_k\leq (C-1)\sum_{k=2}^C a_{k-1}z_k\leq(C-1)x.
\end{equation*}
Then
\begin{align*}
\tilde{R}(z_1,\ldots,z_C) &=\frac{a_1(a_1z_1+y)^2(a_0z_1+x)+a_2(a_1z_1+y)(a_0z_1+x)x+a_3(a_0z_1+x)x^2}{a_0(a_1z_1+y)^3+a_1(a_1z_1+y)^2x+a_2(a_1z_1+y)x^2+a_3x^3}\\
\small
&=\frac{(Cz_1+y)^2(Cz_1+Cx)+(C-1)(Cz_1+y)(Cz_1+Cx)x+(C-1)(C-2)(Cz_1+Cx)x^2}{(Cz_1+y)^3+C(Cz_1+y)^2y+C(C-1)(Cz_1+y)x^2+C(C-1)(C-2)x^3}\\
\normalsize
&\geq\frac{A+(2C^3-C^2)z_1x^2+(3C^2-C)z_1xy+C(C-1)(C-2)z_1x^2}{A+y^3+C^3z_1^2x+2C^2z_1xy+C^2z_1^2y+2Cz_1y^2}\\
&\geq \frac{7y^3+(2C^3-C^2)z_1^2x+(3C^2-C)z_1xy+[C^2(C-1)+C(C-1)(C-2)]z_1x^2}{8y^3+(2C^3-C^2)z_1^2x+(3C^2-C)z_1xy+[C^2(C-1)+C(C-1)^2]z_1x^2}\\
&\geq \min \{\frac{7}{8},\frac{2C-2}{2C-1}\}\\
&=\frac{6}{7}
\end{align*}
where
\begin{align*}
A = Cz_1(Cz_1+y)^2+C(C-1)(Cz_1+y)x^2+Cxy^2.
\end{align*}

The first inequality comes from dropping $C(C-1)(C-2)x^3$ in both the numerator and denominator.  The second inequality follows from the facts that 
$A\geq 7y^3$
and
$C^2z_1^2y+2Cz_1y^2\leq C^2(C-1)z_1^2x+C(C-1)z_1xy+C(C-1)^2z_1x^2,$
since
$y\leq a_1z_1=Cz_1$ and $ y\leq(C-1)x\leq Cx$.
The last inequality follows by the assumption that $C\geq 4$. \Halmos
\end{proof}

\begin{proof}{\underline{\textbf{Proof of Lemma \ref{LemOptimality}}}.}


We derive the optimal condition of the objective function to bound $\lambda_1$ and $\lambda_2$. Recall that our objective function in the case of $C=2$ is
\begin{align}
&\max \ \lambda_1(p(\lambda_1)-c)\mathbb{P}_1+\lambda_2(p(\lambda_2)-c)\mathbb{P}_2\nonumber\\
=&\max \ \frac{2\mu\lambda_1\lambda_2(p(\lambda_1)-c)+2\mu^2\lambda_2(p(\lambda_2)-c)}{\lambda_1\lambda_2+2\mu\lambda_2+2\mu^2}:=f(\lambda_1,\lambda_2)\label{EqTwoN}
\end{align}

Denote $\gamma_i=-p'(\lambda_i)$. Notice that if $\lambda=-ap+b$ is linear, then $\gamma_1=\gamma_2=\frac{1}{a}$. Taking derivative of $f(\lambda_1,\lambda_2)$ w.r.t $\lambda_1$, $\lambda_2$ and set those to zero yields
\begin{align*}
&\frac{\partial f}{\partial \lambda_1}=0 \Rightarrow 2\gamma_1\frac{\lambda_2}{\mu}\left(\frac{\lambda_1}{\mu}\right)^2+2\gamma_1(2\frac{\lambda_2}{\mu}+2)\frac{\lambda_1}{\mu}-(2(p_1-c)(2\frac{\lambda_2}{\mu}+2)-2\frac{\lambda_2}{\mu}(p(\lambda_2)-c))=0\\
&\frac{\partial f}{\partial \lambda_2}=0 \Rightarrow 2\gamma_2(\frac{\lambda_1}{\mu}+2)\left(\frac{\lambda_2}{\mu}\right)^2+4\gamma_2\frac{\lambda_2}{\mu}-2(2(p(\lambda_1)-c)\frac{\lambda_1}{\mu}+2(p(\lambda_2)-c))=0.
\end{align*}
Therefore, the optimal $\frac{\lambda_i^*}{\mu}$'s take the form of
\begin{align}
&\frac{\lambda_1^*}{\mu} = \frac{\sqrt[]{[\gamma_1(\frac{\lambda_2^*}{\mu}+1)]^2+\gamma_1\frac{\lambda_2^*}{\mu}((p(\lambda_1^*)-c)(2\frac{\lambda_2^*}{\mu}+2)-\frac{\lambda_2^*}{\mu}(p(\lambda_2^*)-c))}-\gamma_1(\frac{\lambda_2^*}{\mu}+1)}{\gamma_1\frac{\lambda_2^*}{\mu}}\label{Eq9}\\
&\frac{\lambda_2^*}{\mu} = \frac{\sqrt[]{4\gamma_2^2+4\gamma_2(\frac{\lambda_1^*}{\mu}+2)(2(p(\lambda_1^*)-c)\frac{\lambda_1^*}{\mu}+2(p(\lambda_2^*)-c))}-2\gamma_2}{2\gamma_2(\frac{\lambda_1^*}{\mu}+2)}\label{Eq10}
\end{align}
Notice that by definition, $z_1=\frac{\lambda_1^*}{\mu}\frac{\lambda_2^*}{\mu}$, $z_2=\frac{\lambda_2^*}{\mu}$, and $\beta=\frac{p(\lambda_1^*)-c}{\gamma_1}=\frac{p(\lambda_1^*)-c}{\gamma_2}$. Therefore, we have
\begin{align*}
z_1 &= \frac{\sqrt[]{[\gamma_1(z_2+1)]^2+\gamma_1z_2((p(\lambda_1^*)-c)(2z_2+2)-z_2(p(\lambda_2^*)-c))}-\gamma_1(z_2+1)}{\gamma_1}\\
&\geq \frac{\sqrt[]{[\gamma_1(z_2+1)]^2+\gamma_1z_2(p(\lambda_1^*)-c)(z_2+2)}-\gamma_1(z_2+1)}{\gamma_1}\\
&= \sqrt[]{(z_2+1)^2+\beta z_2(z_2+2)}-(z_2+1)
\end{align*}
and
\begin{align*}
z_2 &= \frac{\sqrt[]{4\gamma_2^2+4\gamma_2(\frac{\lambda_1^*}{\mu}+2)(2(p(\lambda_1^*)-c)\frac{\lambda_1^*}{\mu}+2(p(\lambda_2^*)-c))}-2\gamma_2}{2\gamma_2(\frac{\lambda_1^*}{\mu}+2)}\\
&\leq \frac{\sqrt[]{4\gamma_2(\frac{\lambda_1^*}{\mu}+2)(2(p(\lambda_1^*)-c)\frac{\lambda_1^*}{\mu}+2(p(\lambda_2^*)-c))}}{2\gamma_2(\frac{\lambda_1^*}{\mu}+2)}\\
&=\sqrt[]{\frac{2(p(\lambda_1^*)-c)\frac{\lambda_1^*}{\mu}+2(p(\lambda_2^*)-c)}{\gamma_2(\frac{\lambda_1^*}{\mu}+2)}}\\
&\leq\sqrt[]{\frac{2(p(\lambda_1^*)-c)(\frac{\lambda_1^*}{\mu}+1)}{\gamma_2(\frac{\lambda_1^*}{\mu}+2)}} \leq\sqrt[]{\frac{2(p(\lambda_1^*)-c)}{\gamma_2}} =\sqrt[]{2\beta}.
\end{align*}
\Halmos
\end{proof}

\begin{proof}{\underline{\textbf{Proof of Lemma \ref{Max G}}}.}
We show in this case $G(\beta,z_2)$ is nondecreasing in $z_2$ so that we can plug in the upper bound of $z_2$ to find the maximium of $G(\beta,z_2)$. Letting $A:=\sqrt[]{(1+\beta)z_2^{2}+2(1+\beta)z_2+1}$, then the numerator of $\frac{\partial G(\beta,z_2)}{\partial z_2}$ equals
\begin{equation}\label{Eq11}
\frac{z_2^{3}(3+4\beta+\beta^2)+z_2^{2}(5+3\beta^2+3A+3\beta(2+A))+2z_2(1+\beta^2+A+\beta A)-2\beta A}{A}.
\end{equation}
In the case that $z_2\geq \frac{\sqrt[]{7}-1}{3}$, Equation (\ref{Eq11}) is guaranteed to be non-negative since the coefficient of $\beta A$ equals $3z_2^{2}+2z_2-2$ which is non-negative. Therefore, we can plug in the upper bound of $z_2$ to maximize $G(\beta,z_2)$. By Equation (\ref{z2Opt}) in Lemma \ref{LemOptimality}, we have
\begin{align*}
z_2 \leq \sqrt[]{2\beta}.
\end{align*}

Therefore, 
\begin{equation*}
G(\beta,z_2) \leq \frac{2\beta+\sqrt[]{2\beta}+1-\sqrt[]{(1+\beta)2\beta+2(1+\beta)\sqrt[]{2\beta}+1}}{(3+\beta)2\beta+(2\beta+4)\sqrt[]{2\beta}+2\sqrt[]{2\beta}\sqrt[]{(1+\beta)2\beta+2(1+\beta)\sqrt[]{2\beta}+1}+2}:=h(\beta).
\end{equation*}

Next, we find the maximum value of $h(\beta)$ by looking at the first order condition. Setting $h'(\beta)=0$ yields the following equation,
\begin{equation*}
5\sqrt[]{2\beta}+3\sqrt[]{2}\beta^{5/2}+4\beta^3-4\beta^2B+\beta(2-4B)+2(1+B)-\sqrt[]{2}\beta^{3/2}(2+3B)=0
\end{equation*}
where 
\begin{equation*}
B = \sqrt[]{1+2(\sqrt[]{2}+\sqrt[]{\beta})\sqrt[]{\beta}(1+\beta)}.
\end{equation*}
Let $\theta=\sqrt[]{\beta}$, then we have to solve the following,
\begin{equation}\label{thetaEq}
4\theta^6+3\sqrt[]{2}\theta^5-2\sqrt[]{2}\theta^3+2\theta^2+5\sqrt[]{2}\theta+2=(4\theta^4+4\theta^2+3\sqrt[]{2}\theta-2)\sqrt[]{1+2(\sqrt[]{2}+\theta)\theta(1+\theta^2)}.
\end{equation}
Squaring both sides gives the following polynomial,
\begin{equation*}
16\theta^{12}+56\sqrt[]{2}\theta^{11}+210\theta^{10}+244\sqrt[]{2}\theta^9+316\theta^8+96\sqrt[]{2}\theta^7-18\theta^6-36\sqrt[]{2}\theta^5-36\theta^4-48\sqrt[]{2}\theta^3-66\theta^2-12\sqrt[]{2}\theta=0.
\end{equation*}
The twelve roots to above equation are
\begin{align*}
\theta =& \{-1.59237, -0.951779 \pm 0.164422i, -0.750502 \pm 1.74268i,\\
&-0.547073 \pm 0.940637i, -0.401417, 0, 0.356881 \pm 0.649577i, 0.768987 \}.
\end{align*}
Since $\beta\geq0$, then $\theta=\sqrt[]{\beta}\geq0$. Therefore, only $\theta=0$ and $\theta=0.768987$ can be the only real valued solutions. Notice that $\theta=0$ is not the solution to Equation (\ref{thetaEq}), therefore $\theta^*=0.768987$ is the unique real solution to Equation (\ref{thetaEq}). The corresponding $\beta^*\approx 0.591341$.

Since $h'(0.1)\approx 0.13>0$ and $h'(1)\approx-0.007<0$, then $h(\beta)$ is increasing in $[0,\beta^*]$ and decreasing in $[\beta^*,\infty]$. Therefore, $\beta^*\approx 0.59341$ maximizes $h(\beta)$ where the maximum value is approximately 0.0433.
\Halmos
\end{proof}

%% file: MainPaper.bbl
\begin{thebibliography}{37}
\expandafter\ifx\csname natexlab\endcsname\relax\def\natexlab#1{#1}\fi
\expandafter\ifx\csname url\endcsname\relax
  \def\url#1{{\tt #1}}\fi
\expandafter\ifx\csname urlprefix\endcsname\relax\def\urlprefix{URL }\fi
\expandafter\ifx\csname urlstyle\endcsname\relax
  \expandafter\ifx\csname doi\endcsname\relax
  \def\doi#1{doi:\discretionary{}{}{}#1}\fi \else
  \expandafter\ifx\csname doi\endcsname\relax
  \def\doi{doi:\discretionary{}{}{}\begingroup \urlstyle{rm}\Url}\fi \fi

\bibitem[{Ata and Shneorson(2006)}]{ata2006dynamic}
Ata, Bari{\c{s}}, Shiri Shneorson. 2006.
\newblock Dynamic control of an m/m/1 service system with adjustable arrival
  and service rates.
\newblock {\it Management Science\/} {\bf 52}(11) 1778--1791.

\bibitem[{Balseiro et~al.(2019)Balseiro, Brown, and Chen}]{balseiro2019dynamic}
Balseiro, Santiago, David~B Brown, Chen Chen. 2019.
\newblock Dynamic pricing of relocating resources in large networks.
\newblock {\it Available at SSRN 3313737\/} .

\bibitem[{Banerjee et~al.(2016)Banerjee, Freund, and
  Lykouris}]{banerjee2016pricing}
Banerjee, Siddhartha, Daniel Freund, Thodoris Lykouris. 2016.
\newblock Pricing and optimization in shared vehicle systems: An approximation
  framework.
\newblock {\it arXiv preprint arXiv:1608.06819\/} .

\bibitem[{Banerjee et~al.(2015)Banerjee, Johari, and
  Riquelme}]{banerjee2015pricing}
Banerjee, Siddhartha, Ramesh Johari, Carlos Riquelme. 2015.
\newblock Pricing in ride-sharing platforms: A queueing-theoretic approach.
\newblock {\it Proceedings of the Sixteenth ACM Conference on Economics and
  Computation\/}. ACM, 639--639.

\bibitem[{Bertsekas(2012)}]{bertsekas2005dynamic2}
Bertsekas, Dimitri~P. 2012.
\newblock {\it Dynamic programming and optimal control\/}, vol.~2.
\newblock Athena Scientific.

\bibitem[{Besbes et~al.(2019)Besbes, Elmachtoub, and Sun}]{besbes2018pricing}
Besbes, Omar, Adam~N Elmachtoub, Yunjie Sun. 2019.
\newblock Pricing analytics for rotable spare parts.
\newblock {\it Available at SSRN 3476437\/} .

\bibitem[{Bitran and Mondschein(1997)}]{bitran1997periodic}
Bitran, Gabriel~R, Susana~V Mondschein. 1997.
\newblock Periodic pricing of seasonal products in retailing.
\newblock {\it Management science\/} {\bf 43}(1) 64--79.

\bibitem[{Brumelle(1978)}]{brumelle1978generalization}
Brumelle, Shelby~L. 1978.
\newblock A generalization of erlang's loss system to state dependent arrival
  and service rates.
\newblock {\it Mathematics of Operations Research\/} {\bf 3}(1) 10--16.

\bibitem[{{\c{C}}elik et~al.(2009){\c{C}}elik, Muharremoglu, and
  Savin}]{ccelik2009revenue}
{\c{C}}elik, Sabri, Alp Muharremoglu, Sergei Savin. 2009.
\newblock Revenue management with costly price adjustments.
\newblock {\it Operations research\/} {\bf 57}(5) 1206--1219.

\bibitem[{Chen et~al.(2010)Chen, Wu, and Yao}]{chen2010benefit}
Chen, Hong, Owen~Q Wu, David~D Yao. 2010.
\newblock On the benefit of inventory-based dynamic pricing strategies.
\newblock {\it Production and Operations Management\/} {\bf 19}(3) 249--260.

\bibitem[{Chen et~al.(2015)Chen, Jasin, and Duenyas}]{chen2015real}
Chen, Qi, Stefanus Jasin, Izak Duenyas. 2015.
\newblock Real-time dynamic pricing with minimal and flexible price adjustment.
\newblock {\it Management Science\/} {\bf 62}(8) 2437--2455.

\bibitem[{Chen et~al.(2018)Chen, Farias, and Trichakis}]{chen2018efficacy}
Chen, Yiwei, Vivek~F Farias, Nikolaos~K Trichakis. 2018.
\newblock On the efficacy of static prices for revenue management in the face
  of strategic customers.
\newblock {\it Management Science\/} .

\bibitem[{Chen et~al.(2017)Chen, Levi, and Shi}]{chen2017revenue}
Chen, Yiwei, Retsef Levi, Cong Shi. 2017.
\newblock Revenue management of reusable resources with advanced reservations.
\newblock {\it Production and Operations Management\/} {\bf 26}(5) 836--859.

\bibitem[{Chen et~al.(2006)Chen, Ray, and Song}]{chen2006optimal}
Chen, Youhua, Saibal Ray, Yuyue Song. 2006.
\newblock Optimal pricing and inventory control policy in periodic-review
  systems with fixed ordering cost and lost sales.
\newblock {\it Naval Research Logistics (NRL)\/} {\bf 53}(2) 117--136.

\bibitem[{Cheung et~al.(2017)Cheung, Simchi-Levi, and Wang}]{cheung2017dynamic}
Cheung, Wang~Chi, David Simchi-Levi, He~Wang. 2017.
\newblock Dynamic pricing and demand learning with limited price
  experimentation.
\newblock {\it Operations Research\/} {\bf 65}(6) 1722--1731.

\bibitem[{den Boer(2015)}]{den2015dynamic}
den Boer, Arnoud~V. 2015.
\newblock Dynamic pricing and learning: historical origins, current research,
  and new directions.
\newblock {\it Surveys in operations research and management science\/} {\bf
  20}(1) 1--18.

\bibitem[{Doan et~al.(2019)Doan, Lei, and Shen}]{doan2019pricing}
Doan, Xuan~Vinh, Xiao Lei, Siqian Shen. 2019.
\newblock Pricing of reusable resources under ambiguous distributions of demand
  and service time with emerging applications.
\newblock {\it European Journal of Operational Research\/} .

\bibitem[{Erlang(1917)}]{erlang1917solution}
Erlang, Agner~Krarup. 1917.
\newblock Solution of some problems in the theory of probabilities of
  significance in automatic telephone exchanges.
\newblock {\it Post Office Electrical Engineer's Journal\/} {\bf 10} 189--197.

\bibitem[{Federgruen and Heching(1999)}]{federgruen1999combined}
Federgruen, Awi, Aliza Heching. 1999.
\newblock Combined pricing and inventory control under uncertainty.
\newblock {\it Operations research\/} {\bf 47}(3) 454--475.

\bibitem[{Feng and Gallego(1995)}]{feng1995optimal}
Feng, Youyi, Guillermo Gallego. 1995.
\newblock Optimal starting times for end-of-season sales and optimal stopping
  times for promotional fares.
\newblock {\it Management science\/} {\bf 41}(8) 1371--1391.

\bibitem[{Gallego et~al.(2008)Gallego, Phillips, and
  {\c{S}}ahin}]{gallego2008strategic}
Gallego, Guillermo, Robert Phillips, {\"O}zge {\c{S}}ahin. 2008.
\newblock Strategic management of distressed inventory.
\newblock {\it Production and Operations Management\/} {\bf 17}(4) 402--415.

\bibitem[{Gallego and Van~Ryzin(1994)}]{gallego1994optimal}
Gallego, Guillermo, Garrett Van~Ryzin. 1994.
\newblock Optimal dynamic pricing of inventories with stochastic demand over
  finite horizons.
\newblock {\it Management science\/} {\bf 40}(8) 999--1020.

\bibitem[{Gans and Savin(2007)}]{gans2007pricing}
Gans, Noah, Sergei Savin. 2007.
\newblock Pricing and capacity rationing for rentals with uncertain durations.
\newblock {\it Management Science\/} {\bf 53}(3) 390--407.

\bibitem[{Gong et~al.(2019)Gong, Goyal, Iyengar, Simchi-Levi, Udwani, and
  Wang}]{gong2019online}
Gong, Xiao-Yue, Vineet Goyal, Garud Iyengar, David Simchi-Levi, Rajan Udwani,
  Shuangyu Wang. 2019.
\newblock Online assortment optimization with reusable resources.
\newblock {\it Available at SSRN 3334789\/} .

\bibitem[{Iyengar et~al.(2004)Iyengar, Sigman et~al.}]{iyengar2004exponential}
Iyengar, Garud, Karl Sigman, et~al. 2004.
\newblock Exponential penalty function control of loss networks.
\newblock {\it The Annals of Applied Probability\/} {\bf 14}(4) 1698--1740.

\bibitem[{Kanoria and Qian(2019)}]{kanoria2019near}
Kanoria, Yash, Pengyu Qian. 2019.
\newblock Near optimal control of a ride-hailing platform via mirror
  backpressure.
\newblock {\it arXiv preprint arXiv:1903.02764\/} .

\bibitem[{Kim and Randhawa(2017)}]{kim2017value}
Kim, Jeunghyun, Ramandeep~S Randhawa. 2017.
\newblock The value of dynamic pricing in large queueing systems.
\newblock {\it Operations Research\/} .

\bibitem[{Lei and Jasin(2018)}]{lei2018real}
Lei, Yanzhe, Stefanus Jasin. 2018.
\newblock Real-time dynamic pricing for revenue management with reusable
  resources, advance reservation, and deterministic service time requirements .

\bibitem[{Levi and Radovanovi{\'c}(2010)}]{levi2010provably}
Levi, Retsef, Ana Radovanovi{\'c}. 2010.
\newblock Provably near-optimal lp-based policies for revenue management in
  systems with reusable resources.
\newblock {\it Operations Research\/} {\bf 58}(2) 503--507.

\bibitem[{Ma et~al.(2018)Ma, Simchi-Levi, and Zhao}]{ma2018dynamic}
Ma, Will, David Simchi-Levi, Jinglong Zhao. 2018.
\newblock Dynamic pricing under a static calendar.
\newblock {\it Available at SSRN 3251015\/} .

\bibitem[{Maglaras and Zeevi(2005)}]{maglaras2005pricing}
Maglaras, Constantinos, Assaf Zeevi. 2005.
\newblock Pricing and design of differentiated services: Approximate analysis
  and structural insights.
\newblock {\it Operations Research\/} {\bf 53}(2) 242--262.

\bibitem[{Netessine(2006)}]{netessine2006dynamic}
Netessine, Serguei. 2006.
\newblock Dynamic pricing of inventory/capacity with infrequent price changes.
\newblock {\it European Journal of Operational Research\/} {\bf 174}(1)
  553--580.

\bibitem[{Owen and Simchi-Levi(2018)}]{owen2018price}
Owen, Zachary, David Simchi-Levi. 2018.
\newblock Price and assortment optimization for reusable resources.
\newblock {\it Available at SSRN 3070625\/} .

\bibitem[{Paschalidis and Tsitsiklis(2000)}]{paschalidis2000congestion}
Paschalidis, I~Ch, John~N Tsitsiklis. 2000.
\newblock Congestion-dependent pricing of network services.
\newblock {\it IEEE/ACM transactions on networking\/} {\bf 8}(2) 171--184.

\bibitem[{Rusmevichientong et~al.(2017)Rusmevichientong, Sumida, and
  Topaloglu}]{rusmevichientong2017dynamic}
Rusmevichientong, Paat, Mika Sumida, Huseyin Topaloglu. 2017.
\newblock Dynamic assortment optimization for reusable products with random
  usage durations .

\bibitem[{Waserhole and Jost(2016)}]{waserhole2016pricing}
Waserhole, Ariel, Vincent Jost. 2016.
\newblock Pricing in vehicle sharing systems: Optimization in queuing networks
  with product forms.
\newblock {\it EURO Journal on Transportation and Logistics\/} {\bf 5}(3)
  293--320.

\bibitem[{Yin and Rajaram(2007)}]{yin2007joint}
Yin, Rui, Kumar Rajaram. 2007.
\newblock Joint pricing and inventory control with a markovian demand model.
\newblock {\it European Journal of Operational Research\/} {\bf 182}(1)
  113--126.

\end{thebibliography}
